\documentclass[12pt]{article}
\usepackage{epsf,amsfonts,amssymb,epsfig,amsmath,mathtools,graphics,slashed}
\usepackage{hyperref,hep,graphicx,subfig}
\usepackage{dsfont}
\usepackage{color}

\makeatletter

\newcommand{\Rmnum}[1]{\expandafter\@slowromancap\romannumeral #1@}
\makeatother

\newcommand{\nn}{\notag \\}

\begin{document}

\makeatletter
\renewcommand{\theequation}{\thesection.\arabic{equation}}
\@addtoreset{equation}{section}
\makeatother

\baselineskip 18pt

\begin{titlepage}

\vfill

\begin{flushright}

\end{flushright}

\vfill

\begin{center}
   \baselineskip=16pt
   {\Large\bf Phase transitions of an anisotropic N=4 super Yang-Mills plasma via holography}
  \vskip 1.5cm
  \vskip 1.5cm
      Elliot Banks\\
   \vskip .6cm
         \vskip .6cm
      \begin{small}
      \textit{Blackett Laboratory, 
        Imperial College\\ London, SW7 2AZ, U.K.}
        \end{small}\\

\end{center}

\vfill

\begin{center}
\textbf{Abstract}

\end{center}

\begin{quote}

Black hole solutions of type IIB supergravity were previously found that are dual to N=4 supersymmetric Yang-Mills plasma with an anisotropic spatial deformation. In the zero temperature limit, these black holes approach a Liftshitz like scaling solution in the IR. It was recently shown that these black holes are unstable, and at low temperatures there is a new class of black hole solutions that are thermodynamically preferred. We extend this analysis, by considering consistent truncations of the Kaluza-Klein reduction of IIB supergravity on a five-sphere that preserves multiple scalar and $U(1)$ gauge fields. We show that the previously constructed black holes become unstable at low temperatures, and construct  new classes of exotic black hole solutions.  We study the DC thermo-electric conductivity of these $U(1)$ charged black holes, and find a diverging DC conductivity at zero temperature due to the divergence of the gauge field coupling.
\end{quote}

\vfill

\end{titlepage}
\setcounter{equation}{0}


\section{Introduction}

The AdS/CFT correspondence has proven to be a powerful tool to understand strongly coupled conformal field theories (CFTs), via the construction of black hole solutions in a dual gravitational theory. In order to model realistic systems, it is important that the translational invariance of the CFT is broken, so that there is a mechanism for momentum dissipation within the theory and hence a finite DC linear response. One way to achieve this is to deform the CFT by a spatially dependent source in flat spacetime (for examples see  \cite{Azeyanagi:2009pr,Mateos:2011ix,Mateos:2011tv,Cheng:2014qia,Banks:2015aca,Hartnoll:2012rj,Horowitz:2012ky,Horowitz:2012gs,Donos:2012js,Chesler:2013qla,Ling:2013nxa,Donos:2013eha,
Andrade:2013gsa,Balasubramanian:2013yqa,Donos:2014uba,Gouteraux:2014hca,Donos:2014cya,Jain:2014vka,Donos:2014oha,Donos:2014yya,Kim:2014bza,
Davison:2014lua,Donos:2014gya,Andrade:2014xca,Blake:2015ina,Kiritsis:2015oxa}), which has a number of desireable results. Firstly, these spatially dependent sources allow the construction of novel holographic ground states in the far IR, which provide a way to model a variety of interesting physical states, including insulators, coherent and incoherent metals, and the transitions between them\cite{Donos:2012js,Donos:2013eha,Donos:2014uba,Gouteraux:2014hca,Kiritsis:2015oxa}. Furthermore, many realistic systems appear to be both strongly coupled and anisotropic, including quark gloun plasmas that are formed in heavy ion collisions \cite{Shuryak:2003xe,Shuryak:2004cy}.

For many holographic problems, a phenomenological `bottom-up' approach is used to understand the system. Here, a minimal number of operators are added to the graviatational theory in order to replicate the particular pheneomena one is trying to model. Whilst this often allows for easier computations, the precise holographic dictionary between the bulk graviational theory and the dual CFT is not always fully understood. Whilst computationally more difficult, it is often therefore desireable to take a `top-down' approach, where the graviational dual is constructed from some underlying string or supergravity theory and hence the holographic dictionary is well understood. Top down holographic superconductors have previously shown some unusual features, such as branches of solution that are not thermodynamically perferred and only occur at higher temperatures, a phenomena dubbed `retrograde condensation' or `exotic hair balck holes' \cite{Aprile:2011uq, Donos:2011ut}.

An interesting top down framework to study spatially anisotropic systems in $N = 4$ super Yang-Mills (SYM) theory was developed in \cite{Azeyanagi:2009pr}, and further analysed in various works \cite{Mateos:2011ix,Mateos:2011tv,Cheng:2014qia}. In \cite{Mateos:2011ix,Mateos:2011tv}, black hole solutions of type IIB supergravity that approach $AdS_5$ in the UV were constructed, with a linearly dependent type IIB axion. The solutions were made anisotropic by making the axion a function of only one of the three spatial coordinates. At low temperatures, these black hole solutions approach a $T = 0$ solution from \cite{Azeyanagi:2009pr}, which has a Liftshitz-like scaling in the IR. 

This Liftshitz-like scaling solution is actually unstable at sufficiently low temperatures, and below some critical temperature, the black hole undergoes a phase transition, which spontaneously breaks the global $SO(6)$ symmtery down to $SO(4)\times SO(2)$ \cite{Banks:2015aca}. An explanation for this instability in the zero temperature limit was first presented in \cite{Azeyanagi:2009pr}. There, it was shown that the Kaluza-Klein (KK) spectrum of the five-sphere contains scalar modes, which transform in the $\textbf{20}^\prime$ of $SO(6)$, that saturate the BF bound in an $AdS_5$ background, but violate a similar bound in the Liftshitz scaling solution. In \cite{Banks:2015aca}, it was show that the presence of a scalar field from the KK mode at finite temperatures leads to a phase transition, and a new thermodynamically favoured branch of black hole solutions was constructed. These black holes have interesting properties, such as unusual mean field critical exponents, and undergo a third order phase transition. Whilst the instability from \cite{Banks:2015aca} was due to a single scalar field, this scalar field is only one from the multiplet of twenty, all of which become unstable at the same critical temperature, and it is an interesting question to ask whether the other scalar fields will add to the phase diagram, and what is the true low temperature ground state of the theory.

The work in \cite{Mateos:2011ix,Mateos:2011tv} was generalised in \cite{Cheng:2014qia} to include a finite $U(1)$ gauge field dual to a global $U(1)$ electric field. Whilst most of the physics generalised as expected, it was claimed that the theory has a further global phase transition with the addition of the gauge field. The authors argued that there was an instability in the theory below some critical point, as there are two black hole solutions at the same temperature with different black hole radii, leading to a Hawking-Page transition. It is unclear how the $U(1)$ chemical potential would affect the results of \cite{Banks:2015aca}.

In this paper, we will address both of these questions. We will consider various consistent truncations from type IIB supergravity, first studied in \cite{Cvetic:2000nc}, which preserve multiple scalar fields from the \textbf{20}$'$ multiplet, as well as $U(1)$ gauge fields. The presence of additional scalar fields leads to further branches of black hole solution. However, these solutions form at temperatures above the critical temperature of the phase transition and have a higher free energy - another example of retrograde condensation. Furthermore, by constructing the static normalisable modes of the theory, we will demonstrate that below some critical temperature the solution constructed in \cite{Banks:2015aca} is actually unstable. 

The addition of a finite $U(1)$ chemical potential generalises the results of \cite{Banks:2015aca}, and has many of the same features. In particular, the properties of the phase transition in \cite{Banks:2015aca}, such as the critical exponents and the order of the phase transition, remain unchanged when a chemical potential is switched on. However, for a sufficiently large chemical potential, there is no evidence of retrograde condensation, and the solution appears to be stable all the way down to zero temperatures.  It is important to note, however, that we do not observe the Hawking-Page transition that was observed in \cite{Cheng:2014qia}.

We can then ask questions about the low temperature behaviour of the stable black holes that we have constructed. In particular, by analysisng the DC thermo-electric conductivity of the black holes, using the techniques of \cite{Donos:2015gia,Banks:2015wha}, we find that at zero temperature, the DC thermo-electric conductivity matrix diverges. This is a particular effect of the top down model we have chosen. With the consistent truncation that we have used, at $T = 0$ the gauge coupling in the Lagrangian diverges, and hence the electrical conductivity is infinite.

The rest of the paper proceeds as follows. In section \ref{TDM}, we present the model we will use, discuss the previous work which will be the basis for this study, and introduce our ansatz for the black holes. The black holes are numerically constructed in section \ref{numerics}, and in this section we also dicuss the DC thermo-electric response of the $U(1)$ charged black holes. In the final section we present some conclusions and possible topics of further study.

\section{The model} \label{TDM}
Our starting point is type IIB supergravity. On the field theory side, we are interested in field theories in $3+1$ dimensions, and so we will Kaluza-Klein (KK) reduce the full supergravity theory to $D=5$ using a consistent truncation. By this, we mean that any solution to the equations of motion in the $D = 5$ theory must also be an exact solution to the full ten dimensional theory. In general, obtaining a consistent KK reduction is highly non-trivial in the case of fully backreacted supergravity theories.

A possible starting point is the reduction first obtained in \cite{Cvetic:2000nc}. This truncation preserves 15 $SO(6)$ gauge fields, the type IIB axion and dilaton, and 20 scalar fields that transform in the \textbf{20}$'$ of $SO(6)$, which can be parameterised by a single unimodular symmetric tensor, $T_{ij}$. All twenty scalar fields have mass $m^2 = -4$ \footnote{Throughout the paper we have set the AdS radius, $l = 1$. We have also set $16\pi G = 1$ for the ease of presentation}, which saturates the BF bound in five dimensions, and are dual to operators $\mathcal{O}_\psi$ that have dimension $\Delta = 2$. It is possible to further truncate this theory to all the theories that have been previously studied in \cite{Mateos:2011ix,Mateos:2011tv,Cheng:2014qia,Banks:2015aca}.

Given the computational difficulty in numerically solving the field equations containing this many fields, it will be helpful to further truncate this theory to get something which is more numerically manageable. There are two cases that we will consider - the `neutral' case, where five of the twenty scalar fields are non zero in the consistent truncation with all the gauge fields set to zero, and the `charged' case, which preserves three $U(1)$ gauge fields as well as two scalar fields.

We stress that our use of neutral and charged here simply refers to whether or not the consistent truncation leads to non-zero gauge fields. Unlike in the case of holographic superconductors (e.g \cite{Hartnoll:2008vx}), the phase transitions here will be driven by neutral scalar fields.

\subsection{Neutral case}

We will consider a $D = 5$ gravitational theory coupled to five scalar fields, the dilaton and axion, with a Lagrangian given by
\begin{align}\label{5scalLagrangian}
\mathcal{L} = \sqrt{-g}\left(R - \frac{1}{2}(\partial \vec{\Psi})^2 - V - \frac{1}{2}(\partial \phi)^2 - \frac{1}{2} e^{2\phi}(\partial\chi)^2\right) \, ,
\end{align}
where the potential, $V$, is given by
\begin{equation}
V = -\frac{1}{2}\left((\sum_{i = 1}^6 X_i)^2 - 2\sum_{i = 1}^6 X_i^2\right) \, ,
\end{equation}
$\Psi = (\psi_1,\psi_2,\psi_3,\psi_4,\psi_5)$ and
\begin{equation}
X_i = e^{-\frac{1}{2}\vec{b_i}.\vec{\Psi}} \, .
\end{equation}

The $\vec{b_i}$ satisfy
\begin{equation}
\vec{b_i}.\vec{b_j} = 8\delta_{ij} - \frac{4}{3} \, , \qquad \sum_{i = 1}^6 b_i = 1 \, , \qquad \sum_{i = 1}^6(\vec{u}.\vec{b_i})\vec{b_i} = 8\vec{u} \, ,
\end{equation}
and one convinient choice is \cite{Tran:2001gw}
\begin{align}
\vec{b_1} &= \left(2,\frac{2}{\sqrt{3}},\frac{2}{\sqrt{6}},\frac{2}{\sqrt{10}},\frac{2}{\sqrt{15}}\right) \, , \quad
\vec{b_2} = \left(-2,\frac{2}{\sqrt{3}},\frac{2}{\sqrt{6}},\frac{2}{\sqrt{10}},\frac{2}{\sqrt{15}}\right) \, , \nn
\vec{b_3} &= \left(0,-\frac{4}{\sqrt{3}},\frac{2}{\sqrt{6}},\frac{2}{\sqrt{10}},\frac{2}{\sqrt{15}}\right) \, , \quad
\vec{b_4} = \left(0,0,-\sqrt{6},\frac{2}{\sqrt{10}},\frac{2}{\sqrt{15}}\right) \, , \nn
\vec{b_5} &= \left(0,0,0,-\frac{8}{\sqrt{10}},\frac{2}{\sqrt{15}}\right) \, , \qquad \quad
\vec{b_6} = \left(0,0,0,0,0,-\frac{10}{\sqrt{15}}\right) \, .
\end{align}

This truncation can be obtained from the truncation discussed in the previous section by setting the gauge fields to zero and diagonalising $T_{ij}$\cite{Cvetic:1999xp,Cvetic:1999xx},
\begin{equation}\label{2scal}
T_{ij} = \text{diag}(X_1,X_2,X_3,X_4,X_5,X_6) \, ,\qquad \prod_{i = 1}^6 X_i = 1 \, ,
\end{equation}
and can be further truncated by setting pairs of the elements of (\ref{2scal}) to be equal to each other.  For example, we can set the $X_i$ to be pairwise equal, e.g
\begin{equation}
X_1 = X_2\equiv Y_1 \, , \qquad X_3 = X_4\equiv Y_2 \, , \qquad X_5 = X_6\equiv Y_3 \, ,
\end{equation}
so that
\begin{equation}\label{Trun1}
T_{ij} = \left(Y_1,Y_1,Y_2,Y_2,Y_3,Y_3\right) \, ,
\end{equation}
which leaves us with two remaining scalar fields (since we still have the condition that $T_{ij}$ has unit determinant). The constistency of this truncation was first deduced in \cite{Cvetic:2000eb}. One way to obtain this truncation is to set
\begin{equation}\label{2scalartruncation}
\vec{\Psi} = \left(0, \sqrt{\frac{2}{3}}\psi_1 ,\sqrt{\frac{1}{3}}\psi_1 ,\sqrt{\frac{3}{5}}\psi_2 ,  \sqrt{\frac{2}{5}}\psi_2\right) \, ,
\end{equation}
which explicitly demonstrates that there are two scalar fields remaining.

Furthermore, if we set
\begin{equation}
\vec{\Psi} = \left(0, 0 ,0 ,\sqrt{\frac{3}{5}}\psi ,  \sqrt{\frac{2}{5}}\psi \right) \, ,
\end{equation}
our matrix is now
\begin{equation}
T_{ij} = \left(X,X,X,X,X^{-2},X^{-2}\right) \, ,
\end{equation}
where $X = e^{-\psi/\sqrt{6}}$, and we have the consistent truncation considered in \cite{Banks:2015aca}. Finally, we can set $X = 1$ to recover the Einstein-dilaton-axion theory that was studied by Mateos and Trancanelli \cite{Mateos:2011ix,Mateos:2011tv}.

Note that in order to truncate the five scalar fields to two, we could have equally chosen $\vec{\Psi}$ so that (\ref{Trun1}) was instead $(X_1,X_2,X_3,X_1,X_2,X_3)$, with
\begin{equation}
\vec{\Psi} = \left(\psi_1,\psi_2,-\sqrt{\frac{3}{8}}\psi_1-\sqrt{\frac{1}{8}}\psi_2,\sqrt{\frac{5}{8}}\psi_1-\frac{1}{2}\sqrt{\frac{3}{10}}\psi_2,\frac{2}{\sqrt{5}}\psi_2\right) \, .
\end{equation}
In either case, if both scalar fields acquire a different expectation value, then on uplifting to the full type IIB solution, the global $R$-symmetry will be broken from $SO(6)$ to $SO(2)^3$. Whilst these two examples would appear to be two different solutions in $D = 5$, they are simply part of the moduli space of $D = 10$ IIB solutions, and are physically equivalent.

\subsection{Charged case}\label{Charged model}

We now turn our attention to theories that include gauge fields. We will consider the truncation that preserves two of the scalar fields, three $U(1)$ gauge fields of the maximal $U(1)^3$ subgroup of $SO(6)$, as well as the axion and dilaton, $\phi$ and $\chi$, which was first obtained in \cite{Cvetic:1999xp}. The  Lagrangian for this theory is
\begin{align}\label{QLagrangian}\begin{split}
\sqrt{-g}^{-1}\mathcal{L}
= &R-\frac{1}{2}(\partial\psi_1)^2-\frac{1}{2}(\partial\psi_2)^2 +4\sum_i Y_i^{-1}-\frac{1}{2}(\partial\phi)^2-\frac{1}{2}e^{2\phi}(\partial\chi)^2 \\
&-\frac{1}{4}\sum_iY_i^{-2}(F^i)_{\mu\nu}(F^i)^{\mu\nu}+\frac{1}{4}\epsilon^{\mu\nu\rho\sigma\lambda}F^1_{\mu\nu}F^2_{\rho\sigma}A^3_\lambda \, ,
\end{split}
\end{align}
where the $Y_i$ and $\psi_i$ are the same as in the action described by (\ref{Trun1}), and can be parameterised by
\begin{equation}
Y_i = e^{-\frac{1}{2}\vec{a}_i.\vec{\psi}} \,
\end{equation}
with
\begin{gather}
\vec{a}_1 = (\frac{2}{\sqrt{6}},\sqrt{2}), \qquad  \vec{a}_2 = (\frac{2}{\sqrt{6}},-\sqrt{2}), \qquad  \vec{a}_3= (-\frac{4}{\sqrt{6}},0) \, .
\end{gather}
Although it is not possible to consistently set the gauge fields to zero in the truncation that preserves twenty scalar fields, here we can consistently set the gauge fields to zero to get the consistent truncation which is described by the neutral Lagrangian and (\ref{Trun1}). 

If we turn the dilatonic fields, $\psi_1$ and $\psi_2$, into a single scalar field (the simplest way to does this is to set $\psi_2 = 0$ and let $F^1 = F^2 =F/\sqrt{2}$), our theory reduces the to the truncation studied in \cite{Banks:2015aca} with two additional $U(1)$ gauge fields \footnote{This is Romans $D = 5$ $SU(1) \times U(1)$ gauged supergravity \cite{Lu:1999bw} with an extra dilaton and axion}. Finally, we note that setting both $\psi_1$ and $\psi_2$ to zero and $F^1 = F^2 =F^3 =F/\sqrt{3}$, we now have Einstein-Maxwell-axion-dilaton gravity, which is the model studied in \cite{Cheng:2014qia}.

\subsection{Equations of motion} \label{Solutions}
We now extend the previous analysis of \cite{Azeyanagi:2009pr,Mateos:2011ix,Mateos:2011tv, Banks:2015aca,Cheng:2014qia}, and consider an ansatz of the form
\begin{align}\label{ansatz2}
ds^2 &= \frac{e^{-\frac{1}{2}\phi}}{u^2}\left(-\mathcal{FB}dt^2+\frac{du^2}{\mathcal{F}}+dx^2+dy^2+\mathcal{H}dz^2\right)\,,\nn
\chi &= az, \qquad \phi = \phi(u), \qquad  \psi_i = \psi_i(u), \qquad A^j = b_j(u) dt = b_j dt \, ,
\end{align}
where the $i = 5$ and $j = 0$ for the netural case, and  $i = 2$ and $j = 3$ for the charged case. The UV boundary of our theory is located at $u \rightarrow 0$, whilst there is a black hole horizon at $u_h$. The metric and axion ansatz are the same as in previous works, whilst all other fields only depend on the radial coordinate, $u$, which ensures the ansatz takes the same form as previous work.

The form of the ansatz ensures that the equations of motions will be ODEs, rather than the more technically challenging PDEs. In addition, the form of the gauge potential mean that the Chern-Simons term in (\ref{QLagrangian}) vanishes. The rest of this section proceeds in a similar fashion to section $3$ of \cite{Banks:2015aca}. We therefore only highlight a few key details here, and the refer the reader to the previous paper for full details. Furthermore, whilst this section will refer to the case where the gauge fields are non-zero, the results translate in a straightforward manner if the gauge fields are truncated out.

Substituting this ansatz into the equations of motion, the equation of motion for the axion is trivially satisfied, whilst there are second order equations of motion for the dilaton, gauge and scalar fields in the consistent truncation. In addition, there are four independent components of the Einstein equations, which can be written as equations for $\mathcal{F}',\mathcal{F}'',\mathcal{B}',\mathcal{B}''$. Through appropriate linear combinations of these equations, combined with the equations of motion for the scalar and gauge fields, these four equations reduce to first order equations in $\mathcal{F}$ and $\mathcal{B}$.

The equations of motion are therefore second order in the dilaton, scalar and gauge fields, and first order in $\mathcal{F}$ and $\mathcal{B}$. We can therefore specify the equation of motion by $4 + 2(n_s + n_g)$ integration constants, where $n_s$ and $n_g$ are the number of scalar and gauge fields respectively in the consistent truncation.

At this stage, it is also helpful to note that the ansatz and hence the equations of motion are invariant under the following two scaling symmetries
\begin{align}\label{scsym}
&u\to \lambda u,\quad (t,x,y,z)\to \lambda(t,x,y,z),\quad a\to \lambda^{-1}a, \quad b_i \to \lambda^{-1}b_i;\nn
&t\to \lambda t,\quad {\mathcal B}\to \lambda^{-1/2}{\mathcal B}, \quad b_i \to \lambda^{-1}b_i;
\end{align}
where $\lambda$ is a constant.

\subsection{Boundary conditions}
We will now disucss the boundary conditions for our theory, and hence derive the expansions for the functions near the boundary $u \rightarrow \infty$, the UV, and near the black hole horizon at $u_h$, the IR. 

First, we consider the UV Expansion. We require that our solution asymptotically approaches $AdS_5$ with an axionic field that is deformed by strength $a$ in the $z$ direction. In order to have the correct falloff, we require $\phi \rightarrow 0$, and $\psi_i \rightarrow 0$. Furthermore, we require that the three gauge fields tend to constant values at the boundary, corresponding to switching on a chemical potential. In order to make a connection with \cite{Cheng:2014qia}, we will take the three gauge fields to have the same chemical potential, $\mu$.

By imposing these boundary conditions on the solutions, and solving the equations of motion order by order, we see that the solution has an asymptotic expansion:
\begin{align}\label{uvexp}
{\phi} &= -\frac{a^2 u^2}{4}+\dots\,,\nn
\mathcal{F} &= 1+\frac{11 a^2 u^2}{24}+u^4\mathcal{F}_4 +u^4\log u \frac{7a^4}{12}  +\dots\,,\nn
\mathcal{B}&= 1-\frac{11 a^2 u^2}{24}+u^4\mathcal{B}_4-u^4\log u  \frac{7a^4}{12} +\dots\,,\nn
\psi_i &= \langle\psi\rangle_iu^2+\dots\, , \nn
b_i &= \mu+\frac{1}{2}\rho_i u^2+\dots\, ,
\end{align}
where $\langle\psi\rangle_i$ corresponds to the VEV of the operator dual to $\psi_i$ and $\rho_i$ is the electric current for the gauge field $A_i$. We have set terms proportional to $u^2\log u$ to zero in the expansions of $\psi_i$, which mean that there is no source term for the operator dual to $\psi_i$, and we have used the second scaling symmetry from (\ref{scsym}) to set $\mathcal{B}$ to 1. We can see that the expansion is determined by $4 +n_s + n_g$ terms. The $\log$ terms in the expansion indicate that there is a conformal anomaly in our theory, which introduces an additional dynamical scale. However, for the present purposes it will suffice to hold this scale to be fixed to unity throughout.

We now consider the IR expansion, and demand that the black hole has a regular event horizon at $u_h$, which requires the gauge fields and $\mathcal{F}$ vanish on the horizon. We find that the leading order expansion about $u_h$ for the fields is given by
\begin{align}\label{irexp}
\mathcal{F} &= -\frac{4\pi T}{\sqrt{\mathcal{B}_h}}\left(u-u_h\right) +\dots\,, \qquad \mathcal{B} = \mathcal{B}_h+\dots\, , \qquad \phi = \phi_h+\dots\,,\nn
{\psi_i} &=\psi_{ih}+\dots \, , \qquad b_i = a_{ih}(u-u_h) +\dots, \, ,
\end{align}
where $T$ is the Hawking temperature of the black hole, which can be expressed in terms of $u_h$ and the other free parameters in the expanion. There are therefore $3 +n_s + n_g$  parameters in the IR expansion of the fields.

Combining these two results, we see that our system is determined by $7 + n_s + n_g$ constraints, which, after applying the remaining symmetry from (\ref{scsym}) gives us $6 + 2(n_s + n_g)$  constants of integration. Recalling that the order of our equations is $4 + 2(n_s + n_g)$, the black holes are specified by a two parameter family of solutions. Throughout the remainder of the paper, we will use the grand canonical ensemble to describe the system and so these parameters will be $T/a$ and $\mu/a$. In the `neutral case' we have a one parameter family of solutions specified by $T/a$.

\section{Numerical construction of the black holes}\label{numerics}
We can now solve the equations of motion numerically in order to construct black hole solutions. As discussed previously, there are two cases that we will consider - the `neutral' case where we have set the chemical potential to zero, and `charged' case, where we source all gauge fields with the same, constant chemical potential, $\mu$. Since our boundary conditions ensure that there is no source to the scalar fields, $\psi_i$, any new branch of solution will correspond to a phase transition by spontaneous symmetry breaking, since the dual operator to the field, $\mathcal{O}_\psi$ will have a finite expectation value but no source.

For both cases, our numerical method has been to set the anisotropic strength, $a$, to 1, and use a shooting method to solve the equations. This means that we numerically integrate the solution from both the black hole horizon and the UV boundary, and match at some midpoint between the two boundaries. The Smarr relation
\begin{equation}
E-Ts+\sum_i\rho_i\mu=-T^{xx} \, ,
\end{equation}
where $E = T^{tt}$, $T^{\mu\nu}$ is the stress energy tensor of our black hole and $s$ is the entropy density given by
\begin{equation}\label{ent}
s = 4\pi \frac{e^{-\frac{5}{4}\phi_h}}{u_h^3} \, ,
\end{equation}
provides a useful check of the numerics. This relation can be verified using the method outlined in Appendix A of \cite{Banks:2015aca}.

Since there are several branches of black hole solution that have been numerically constructed, in what follows we will refer to the black hole solutions from \cite{Mateos:2011ix,Mateos:2011tv} as the `Mateos-Trancanelli' solution, the black hole from \cite{Cheng:2014qia} as the `Cheng-Ge-Sin' (CGS) solution, and the new branch of black hole solution constructed in \cite{Banks:2015aca} as the `1 scalar' solution.

\subsection{Neutral case}\label{NCase}

\begin{figure}
\centering
\includegraphics[scale = 0.6]{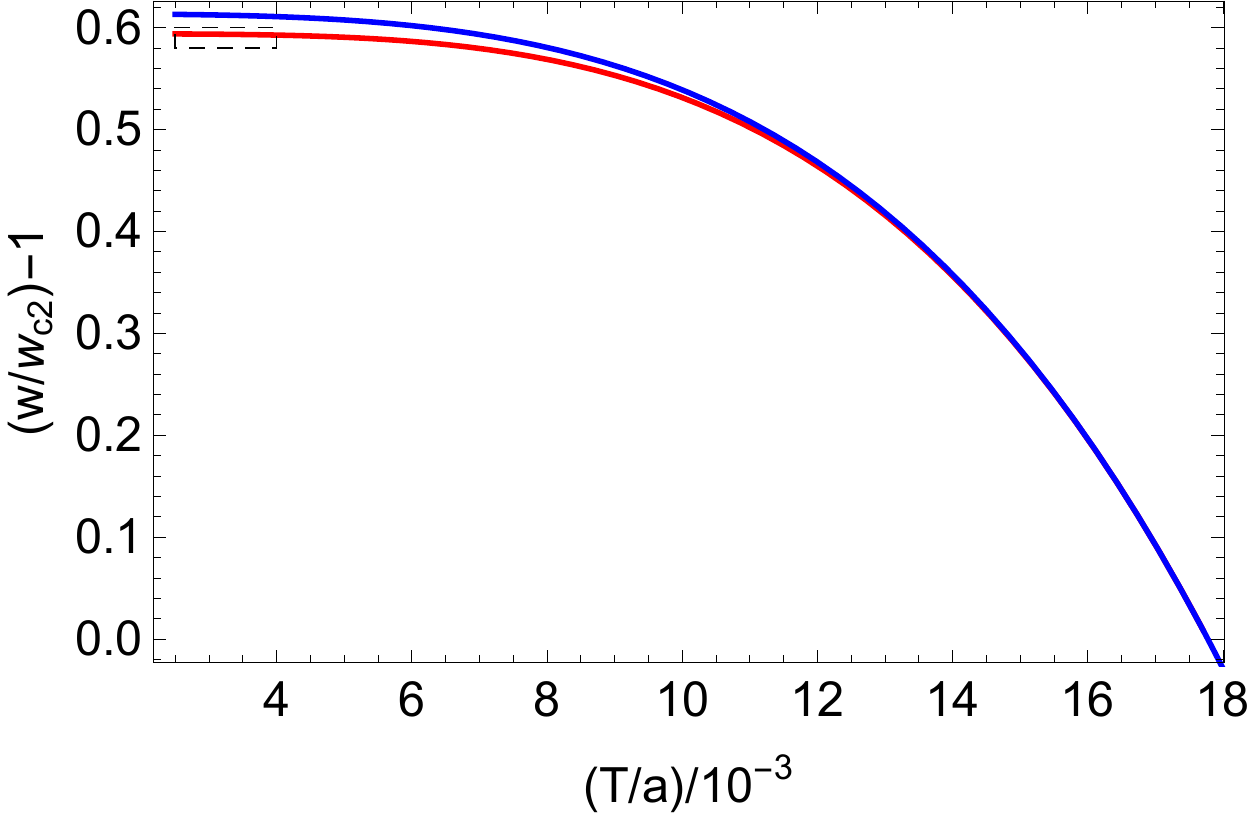} \, \\
\includegraphics[scale = 0.55]{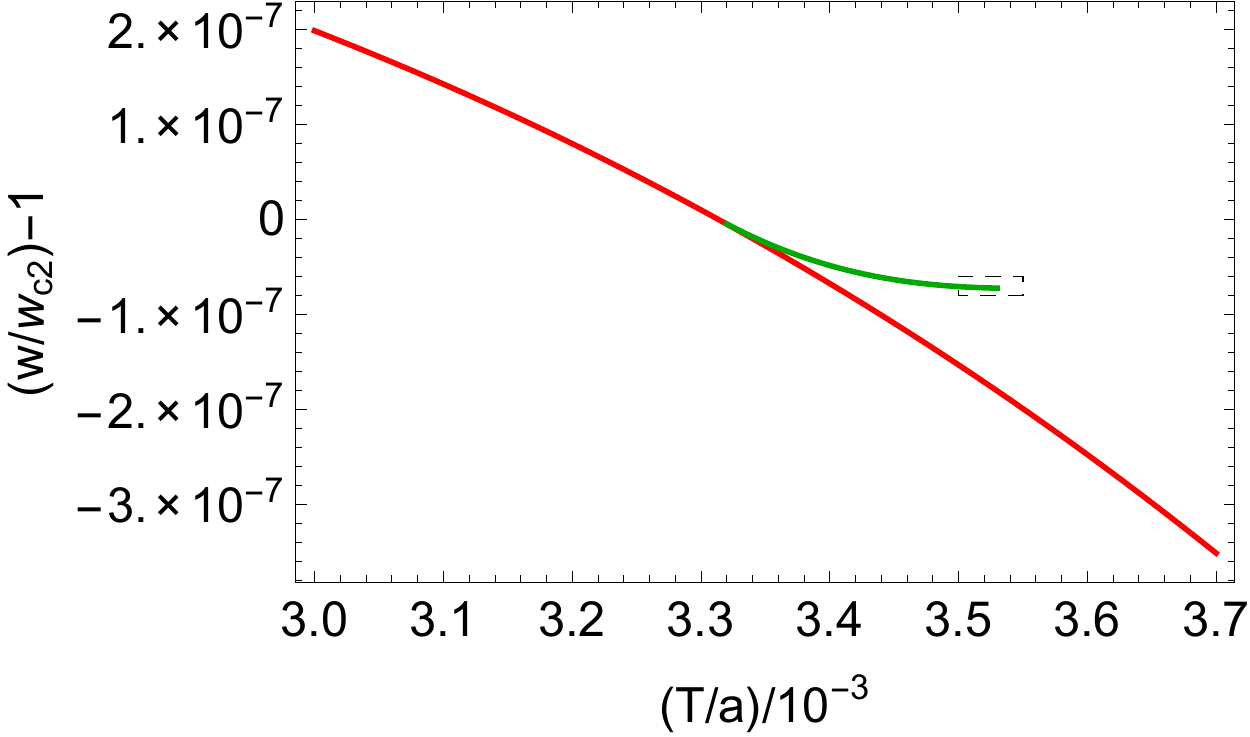} \qquad
\includegraphics[scale = 0.55]{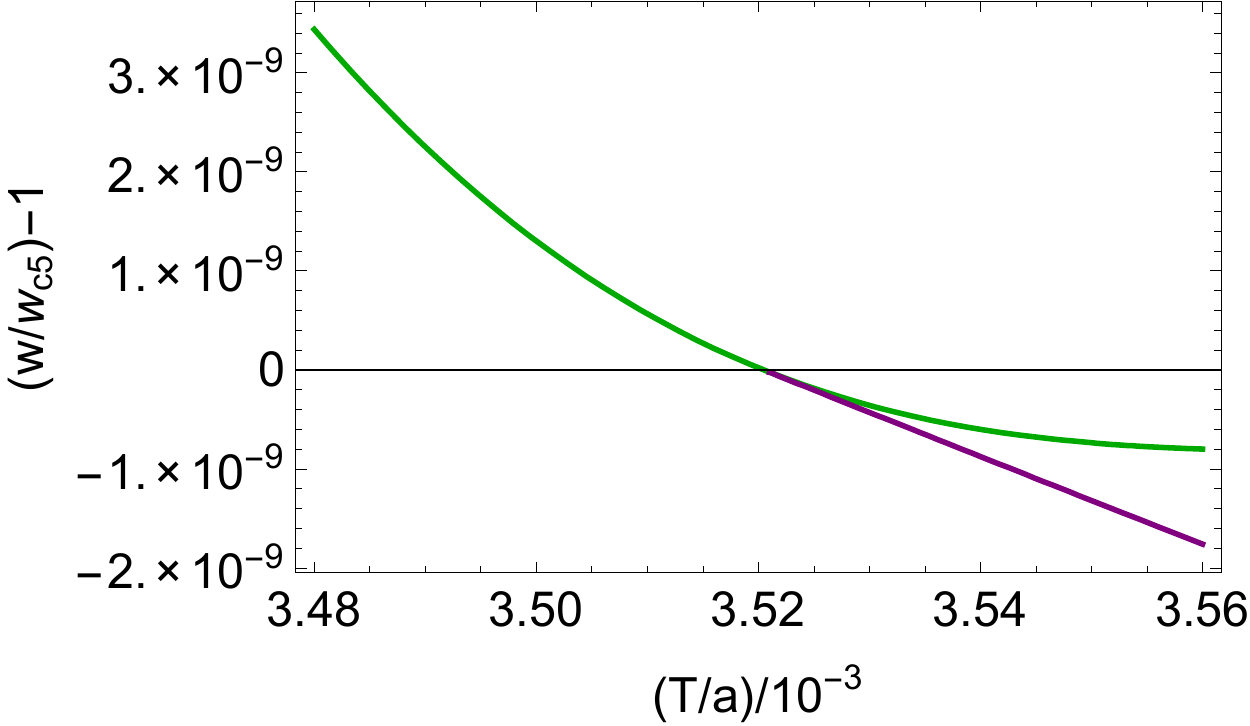} 
\caption{Plot showing the free energy of the black hole solutions in the netural case, scaled in relation to the free energy at the critical temperature of the particular phase transition. The dashed boxes in the top plot shows the region that has been magnified in the bottom left plot, whilst the dashed box in the bottom left chart has been magnified in the bottom right plot. The blue line is the Mateos-Trancanelli solution, the red line is the 1 scalar solution from \cite{Banks:2015aca}, whilst the green and purple lines indicate new branches of black hole solution, that only form at a temperature above the critical temperature of the phase transition. Furthermore, these solutions have a higher energy than the 1 scalar solution and are not thermodynamically preferred, but instead correspond to retrograde condensation.
}
\label{NeutralSolution}
\end{figure}

\begin{figure}
\centering
\includegraphics[scale = 0.75]{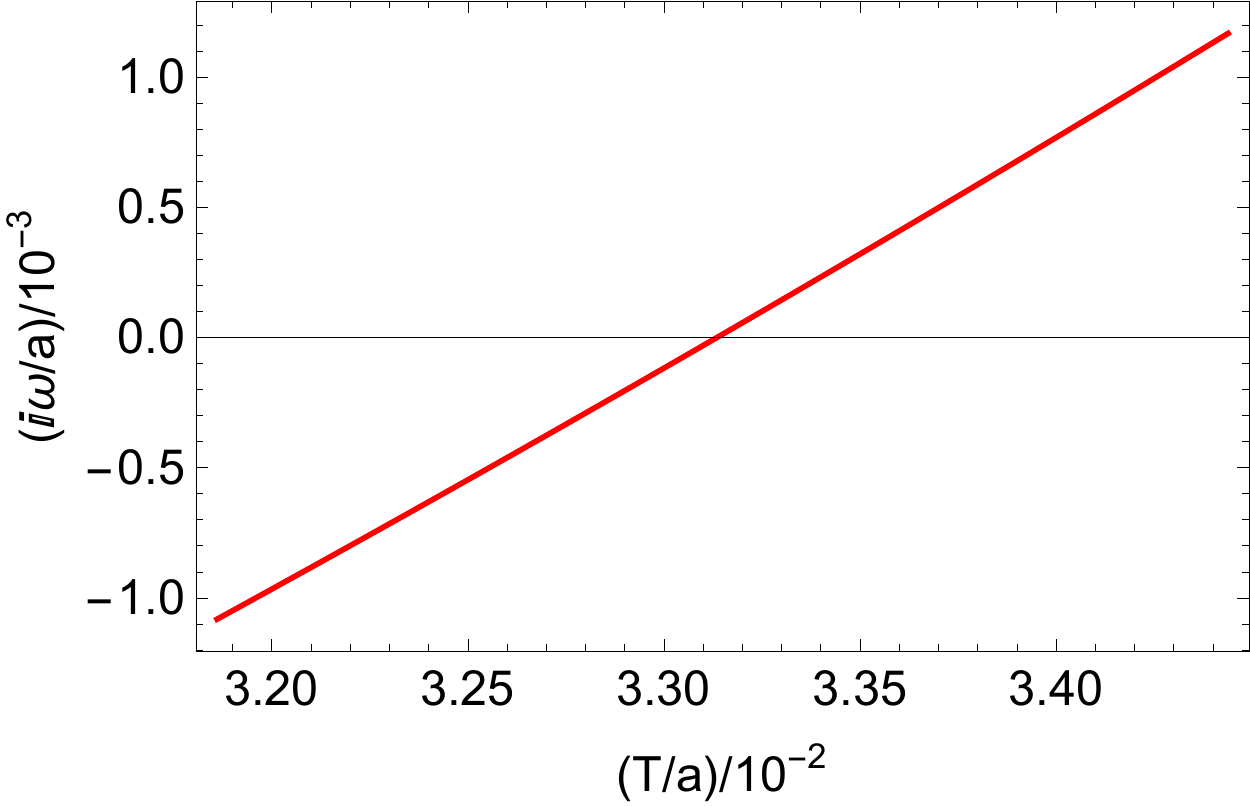} 
\caption{Frequency of normalisable mode versus temperature, for the 1 scalar branch of solution. Below the critical temperature, $T_{c2}/a$, any perturbation of the form $e^{-i\omega t}$ will grow exponentially in time, and will cause an instability. The quasinormal mode frequency appears to be purely imaginary and hence is a purely decaying mode.}
\label{wplot}
\end{figure}

Recall that in \cite{Banks:2015aca}, it was shown that the Mateos-Trancanelli solution (denoted by the blue line in figure \ref{NeutralSolution}) is unstable below a critical temperature, $T_{c1}/a \sim 1.8 \times 10^{-2}$, and a new branch of black hole solution is formed which is thermodynamically preferred (denoted by the red line). This instability is driven by the condensation of a single scalar field that transforms in the $\textbf{20}^\prime$ of $SO(6)$ and has a mass $m^2 = -4$. Here, we extend this analysis by including five of the scalar fields from the multiplet, all of which all have $m^2 = -4$, and find further branches of black hole solution.

The additional scalar fields in the Lagrangian (\ref{5scalLagrangian}) lead to further branches of black hole solution, which are show in figure \ref{NeutralSolution}. At a critical temperature $T_{c2}/a \sim 3.3 \times 10^{-3}$, a new branch of solution (the green line) appears from the 1 scalar solution, whilst further along this branch there is another phase transition that occurs at $T_{c5}/a$ and leads to a further branch of solutions (the purple line). The green branch of solution has two independent scalar fields (and so is also a solution to the consistent truncation described by (\ref{Trun1})), whilst the purple branch of solution has three independent scalar fields (and so is a solution to equations of motion with two pairs of $X_i$ in \ref{2scal} set equal). On uplifting to the full $D = 10$ theory we see that the global $R$-symmetry of the solution is first broken from $SO(4) \times SO(2)$ to $U(1)^3$ and then to $U(1)^2$ along the branches.

Although our Lagrangian contains five scalar fields, it appears that in our solutions there are only a maximum of three fields that are actually independent. This can be seen by analysing the linear perturbation of the five scalar fields around the green branch of solutions. It is only possible to get consistent equations of motion when one of the scalar fields is perturbed in this background, and so only one additional scalar field will condense at $T_{c5}/a$.

Unlike in \cite{Banks:2015aca}, when the Mateos-Trancanelli solution undergoes a phase transtiion to the 1 scalar solution, the branches of solution constructed here only exist at temperature $T \ge T_{c2}/a$, and are not thermodynamically perferred. Furthermore, we find no evidence that the branch will turn back to lower temperatures (which would indicate a first order transition). This is an example of retrograde condensation and has been observed in top down models for holographic superconductors \cite{Aprile:2011uq, Donos:2011ut}. 

In \cite{Donos:2011ut}, black brane solutions were constructed that also displayed this retrograde condensation. In this example, the black hole solution itself was unstable below the critical point, with a nakedly singular solution at $T = 0$.  It is an interesting question to therefore ask if the geometry here is unstable below $T_{c2}/a$. To do this, we introduce a perturbation of the form $e^{-i\omega t}\delta \psi_2$ and impose infalling boundary condition. We then ask what value of $\omega$ do we get a normalisable fall-off, ie at what value of $\omega$ can we get a VEV for the scalar field without a source. A plot of $i\omega$ for values of $T/a$ is shown in figure \ref{wplot}. We find that for $T > T_{c2}$, $i\omega$ is positive, and so any perturbation of the scalar field will decay over time. However, below $T_{c2}$, $i\omega$ changes sign, which means that a perturbation of the scalar field will grow exponentially with time. We therefore conclude that the theory becomes unstable below the critical point, $T_{c2}/a$.

\subsection{Charged case}
When we add a single, finite $U(1)$ chemical potential, we observe interesting results which highlight the competing effects that geometry and charge have on phase transitions in top down holographic models. Our starting point is the CGS solution from \cite{Cheng:2014qia}, which is the charged analogue of the Mateos Trancanelli solution. In their paper, the authors found that the CGS solution  has some minimum temperature and the black hole undergoes a Hawking-Page style phase transition below this point. Here, we do not find this, but instead find that the black hole approaches an extermal black hole in the low temperature limit, which we discuss in further detail in Appendix \ref{Extremal}. However, as we now explain, the CGS solution is unstable below a critical temperature, $T_{c1}/a$

\begin{figure}
\centering
\includegraphics[scale = 0.55]{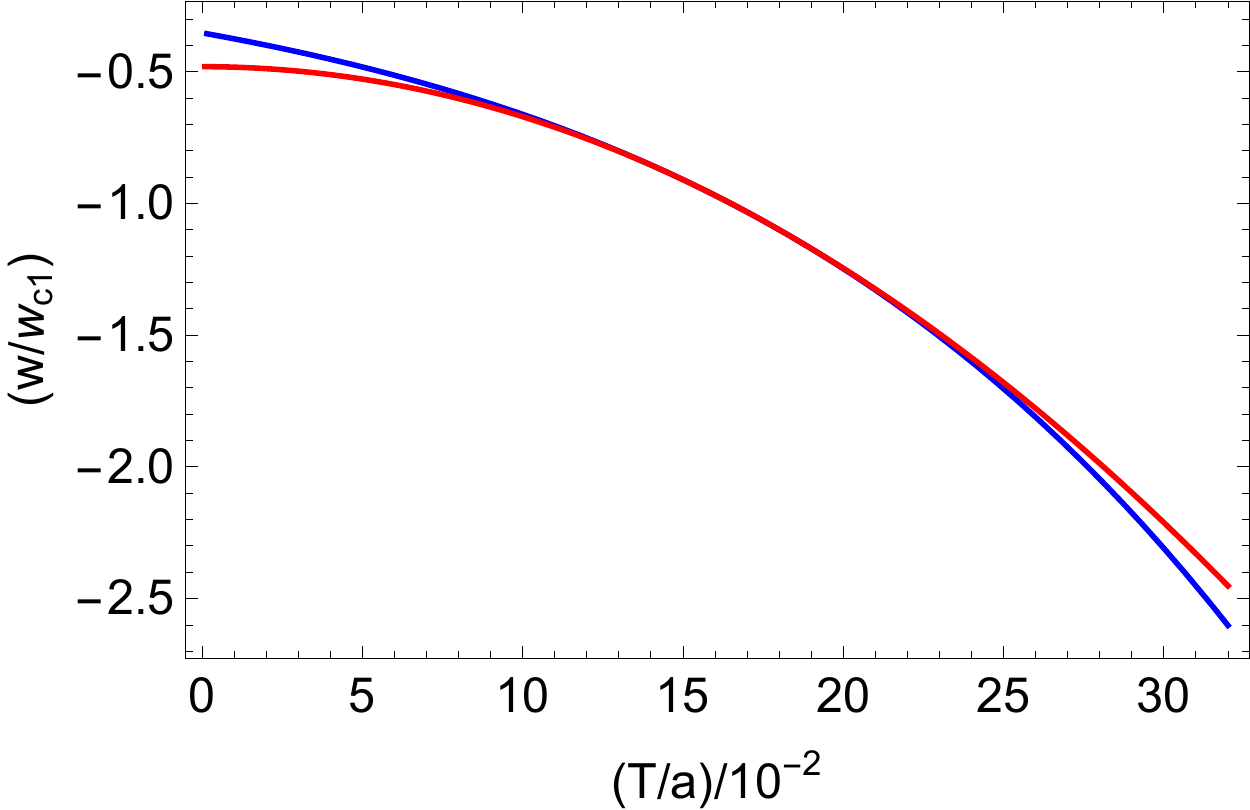} \qquad
\includegraphics[scale = 0.55]{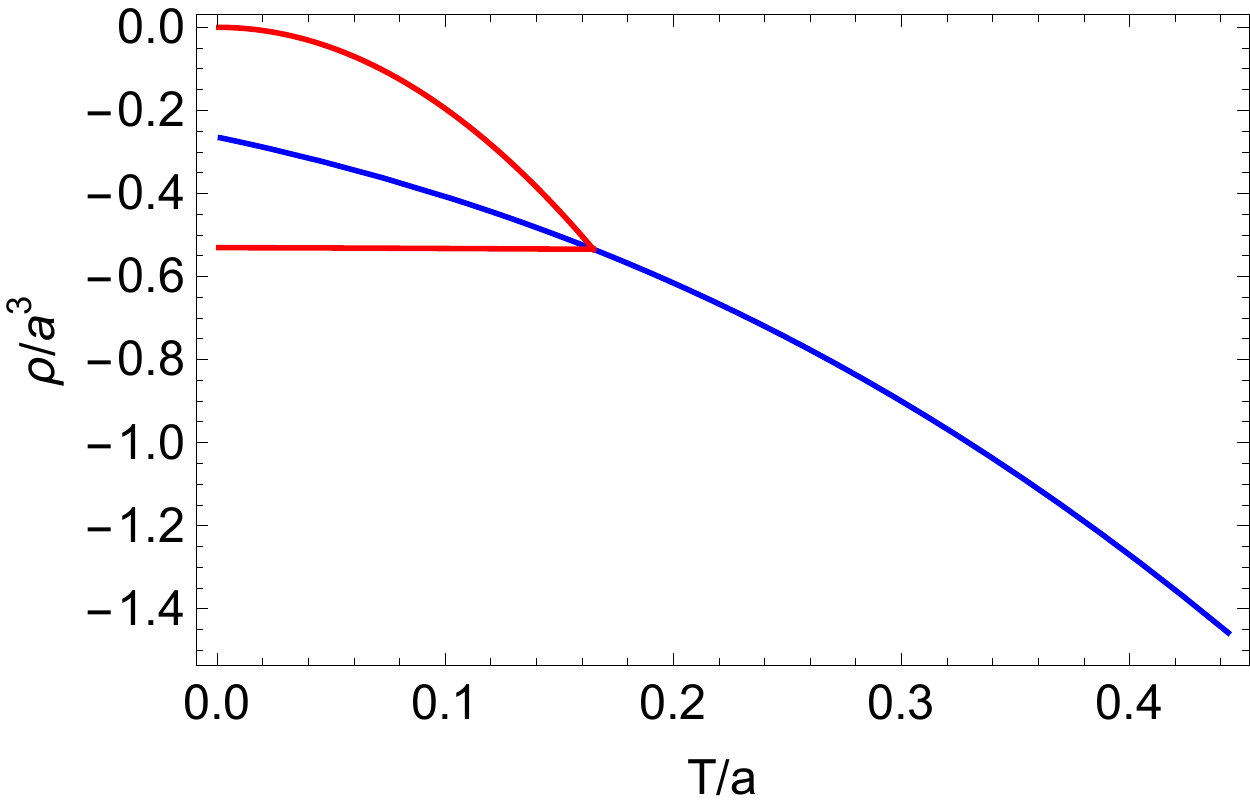} 
\caption{Plot showing the charged branches of black hole solutions when $\mu/a = 1$. The blue line is the CGS soution from  \cite{Cheng:2014qia}, whilst the red line is the branch of solution analagous to that found in \cite{Banks:2015aca}. The left plot shows the free energies of the solutions, and there is a branch of solution that forms at temperatures above and below the phase transition. However, only the lower temperature new solution is thermodynamically preferred. The right hand plot shows the electric charge density when the chemical potential is turned on, for the CGS solution and the new branch of solution. The solution has only been shown below the critical temperature for the ease of presentation.  Below the critical temperature, the new branch of solution has two different electric charge densities due to the symmetry breaking.
}
\label{ChargedSolutions}
\end{figure}

The phase transition observed in \cite{Banks:2015aca} is also present when the chemical potential is turned on - in this case, the CGS solution becomes unstable below $T_{c1}/a$ (which now depends on $\mu/a$), and the plasma undergoes a third order phase transition (with the same critical exponents as \cite{Banks:2015aca}) to a new branch of thermodynamically preferred black hole solutions. As before, there is also an unphysical branch of solutions that forms above the critical temperature, with a higher free energy than the background solution, corresponding to an exotic hairy black hole. The free energy for these solutions with $\mu/a = 1$ is shown in the left hand plot of figure \ref{ChargedSolutions}, where the blue line is the CGS solution and the red line is the new branch.

As one would expect from the discussion in (\ref{Charged model}), the multiple gauge fields in our theory can now be explicitly seen in this new branch of solution, due to the breaking of the symmetry in the $D = 10$ supergravity theory. In left plot of  figure \ref{ChargedSolutions}, we plot the electric charge density of the solutions against temperature. The blue line is the charge density for the CGS solution, whilst the two red lines are the two different charge densities in this new branch of solution below the critical temperature.

\begin{figure}
\centering
\includegraphics[scale = 0.55]{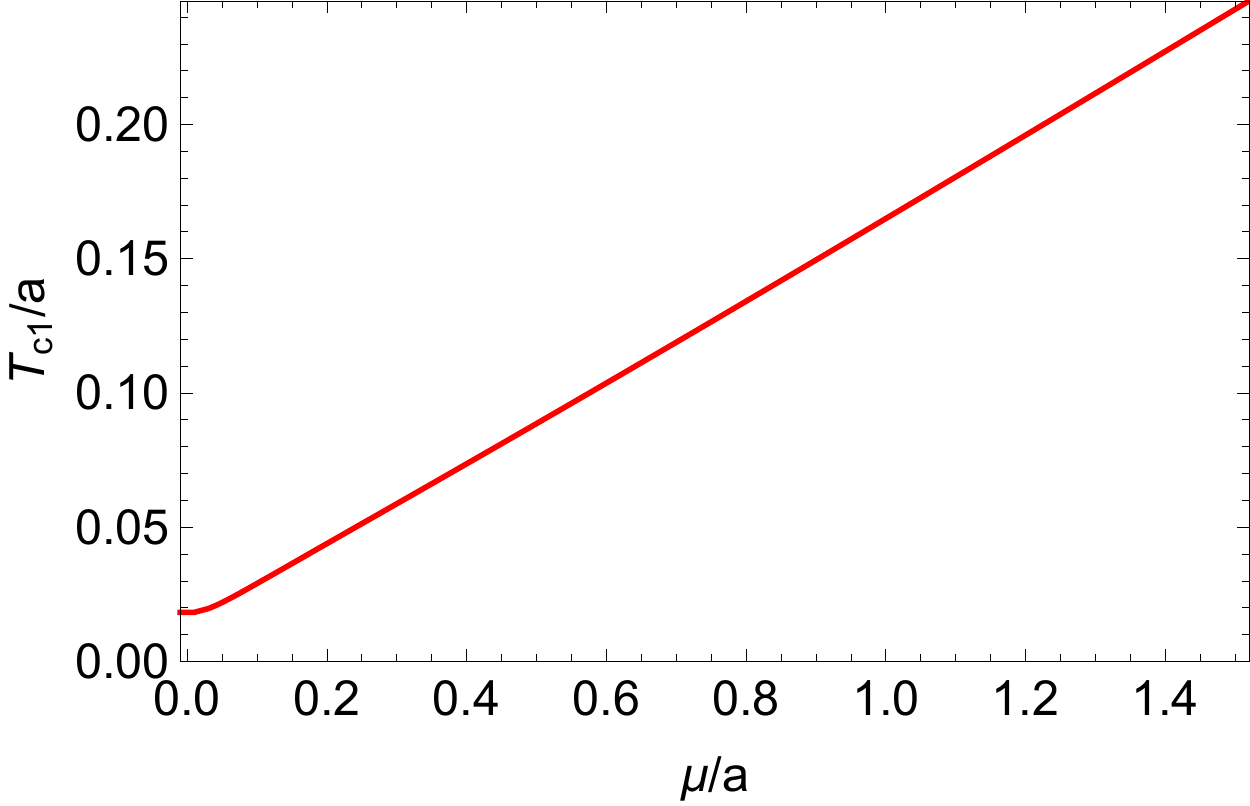} \qquad
\includegraphics[scale = 0.55]{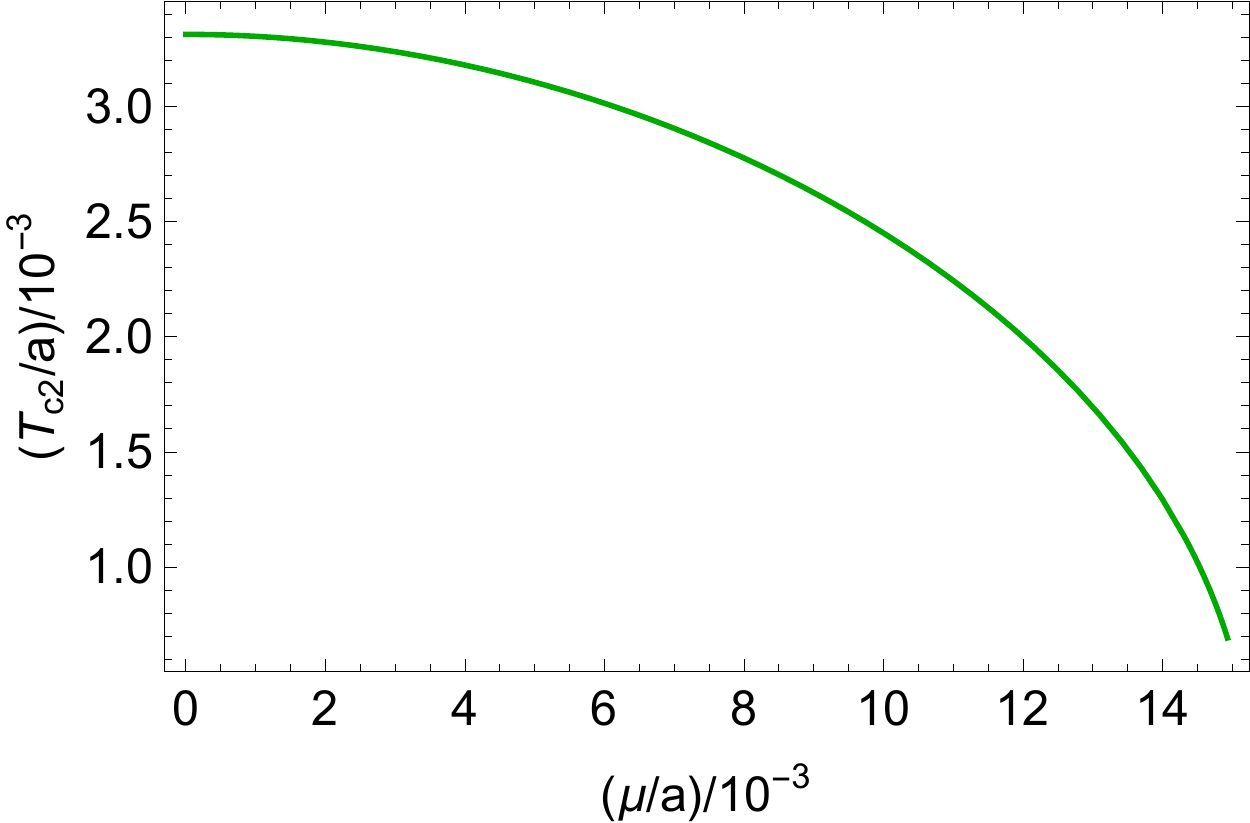} 
\caption{Plot showing critical temperatures, $T_{c1}/a$ (left) and $T_{c2}/a$ (right) as a function of chemical potential, $\mu/a$. The phase transition seen in \cite{Banks:2015aca} occurs for all values of chemical potential. However, the instability that discussed in section \ref{NCase} is only seen for $\mu/a < 0.015$. The symmetry of the gauge field ensures these results are symmetric if we take $\mu \rightarrow -\mu$.
}
\label{MuPlot}
\end{figure}

We find that the critical temperature of this phase transition, $T_{c1}/a$, increases as we increase the chemical potential, and so the phase transition found in \cite{Banks:2015aca} is at the lowest critical temperature in our class of black holes. We have checked the relationship between critical temperature and chemical potential up to $\mu/a \sim 1.5$, and have no reason to suspect that this will change at higher chemical potentials. The chemical potential dependence of the critical temperature for this phase transition is shown in the red plot of figure \ref{MuPlot}.

However, for the condensation of a second scalar field (analagous to the second plot in figure \ref{NeutralSolution}), there is a different story. Although this phase transition is seen for small $\mu/a$, when $\mu/a \ge \mu_c/a \sim 0.015$, this lower temperature phase transition is no longer seen. The green plot in figure \ref{MuPlot} shows the effect of increasing $\mu/a$ has on this critical temperature for the phase transition, $T_{c2}/a$. The critical temperature decreases until at $\mu = \mu_c$ the critical temperature is zero, and the phase transition no longer occurs. Interestingly, this means that although our black holes appear to be unstable to scalar field perturbations at low temperatures in the neutral case, the solutions can be `saved' by turning on a sufficiently large chemical potential.

\subsection{Charged solution thermo-electric DC conductivity}\label{Charged}
We now change tack, and study some of the properties of the new charged branch of solutions\footnote{For all the calculations in this section, unless otherwise stated, we have set $\mu/a = 0.02 $, to ensure that the solution is stable at low temperatures, as per the discussion in the previous section.}. As the solution is cooled down to low temperatures, thermodynamic quantities such as entropy begin to show temperature dependent scaling. This indicates that, unlike in the CGS solution, the zero temperature black hole solution is not extremal and instead the black hole reaches zero temperature as $u_h \rightarrow \infty$.

Since the translational invariance of the theory has been broken in the $z$ direction, it is interesting to determine the DC transport coefficients of the phase in this direction. Since there are two distinct gauge fields, we anticipate that there should be two different electric conductivities. This leads to a $3 \times 3$ thermo-electric conductivity matrix in the form
\begin{align}\label{bigform}
\left(
\begin{array}{c}
J_1\\J_2\\Q
\end{array}
\right)=
\left(\begin{array}{ccc}
\sigma_{11} & \sigma_{12} & \alpha_1 T \\
\sigma_{21} &\sigma_{22} & \alpha_2 T   \\
\bar\alpha_1 T & \bar \alpha_2 T& \kappa \\
\end{array}\right)
\left(
\begin{array}{c}
E_1\\E_2\\-(\nabla T)/T
\end{array}
\right)\,,
\end{align}
where the $J$'s are the electric current densities from the two gauge fields, the $Q$ is the heat current density, while $E_i$ and $\nabla T$ are the applied electric fields and thermal gradients. The symmetries of our ansatz imply that $\sigma_{12} = \sigma_{21}$, and $\bar\alpha_i = \alpha_i$.

In order to determine this DC thermo-electirc conductivity matrix, we adopt the method of \cite{Donos:2014cya,Donos:2014yya} and consider a linearised perturbation of the form 
\begin{align}\label{pertansatz}
A^a_{z}&=-\delta f_1^a(u)t+\delta a^a_z(u)\,,\nn
g_{tz}&=t\delta f_2(r)+\delta g_tz(u)\,,\nn
g_{uz}&=\delta g_{uz}(u)\,,\nn
\delta \phi &= \delta \phi(u) \, ,\nn
\delta \chi &= \delta \chi(u) \, ,
 \end{align}
where $a$ is a label for the gauge fields, and $\delta f_1$ and  $\delta f_2$ are related to the sources for the electric and heat current respetively.  The matrix (\ref{bigform}) can then be determined by evaluating the heat and electric current on an appropriate radial hypersurface. The details of this calculation in a general setting, which extends the results of \cite{Donos:2015gia,Banks:2015wha}, are described in appendix \ref{DC}.

\begin{figure}
\centering
\includegraphics[scale = 0.55]{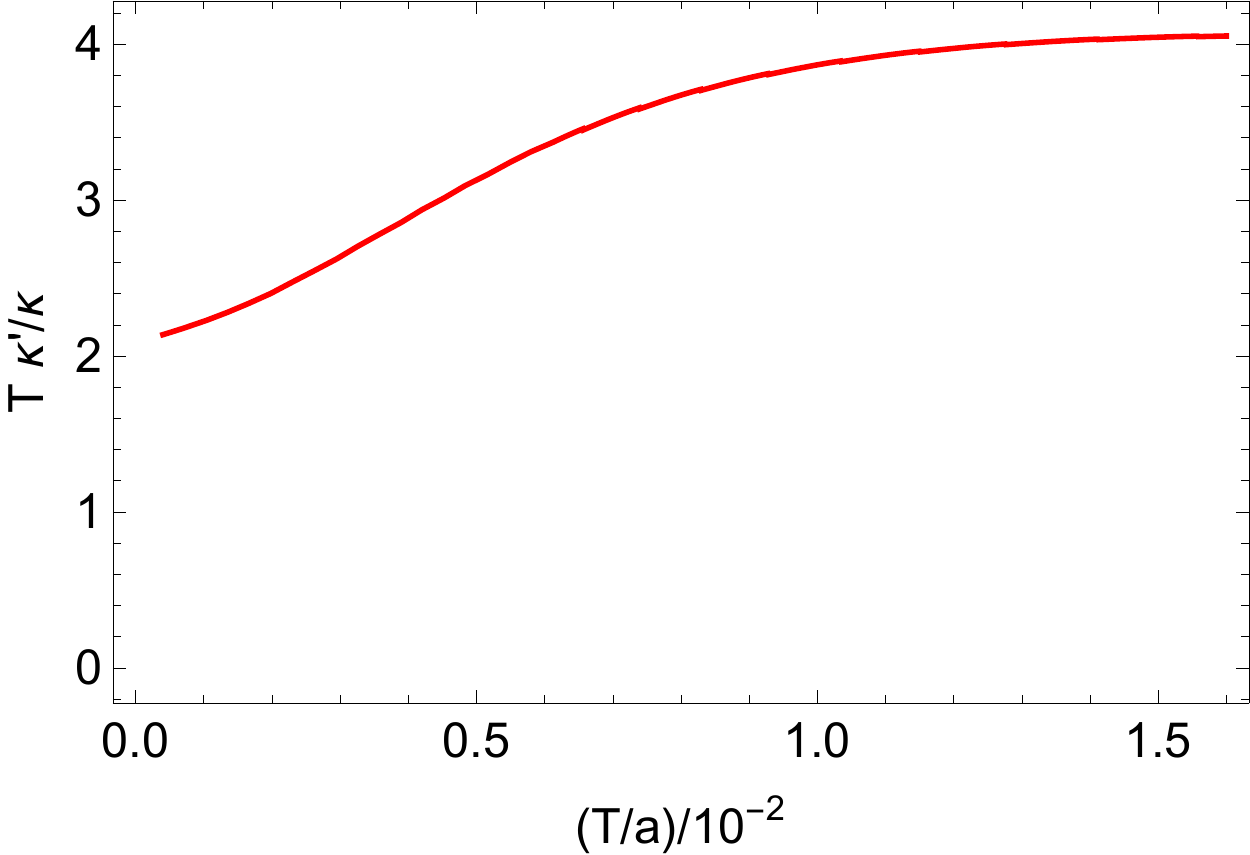} \qquad 
\includegraphics[scale = 0.56]{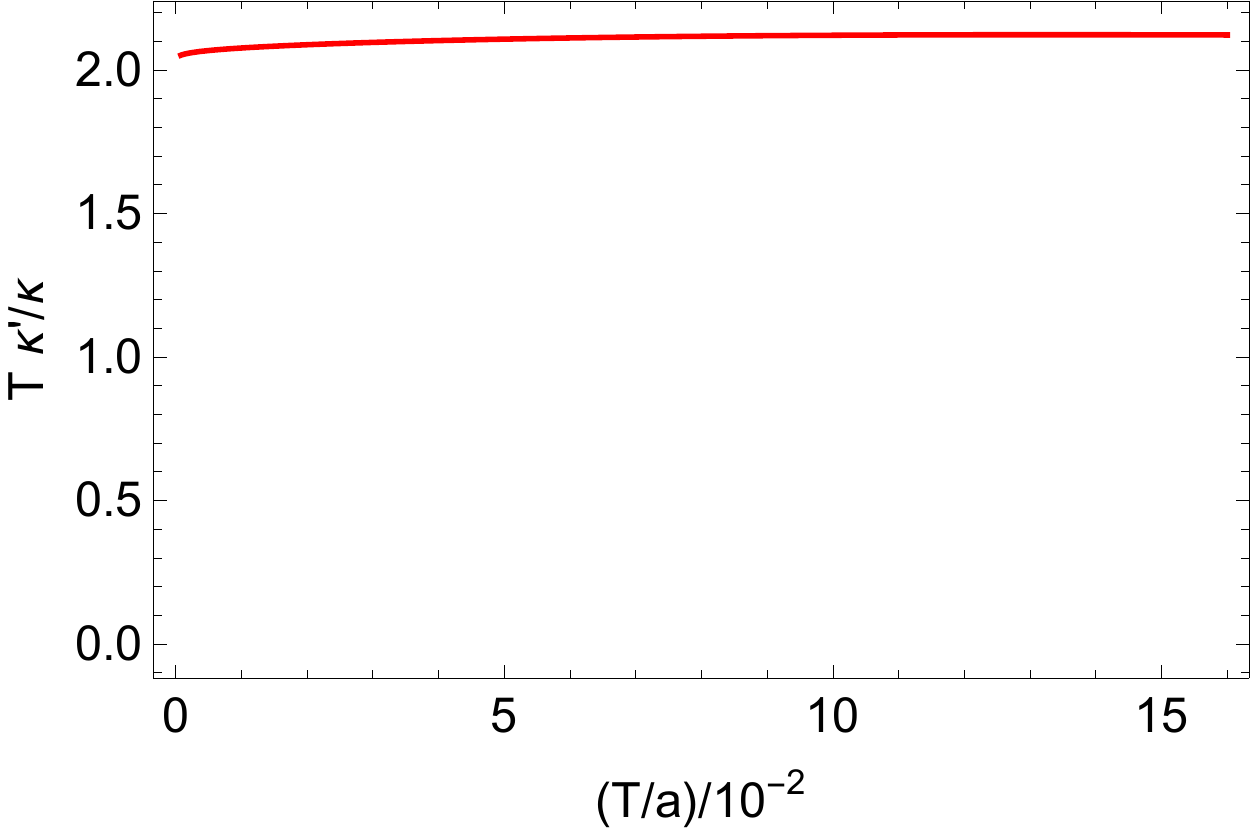}
\caption{Plot showing the scaling of the DC thermal conductivity, $\bar \kappa$ against temperature, $T/a$, at $\mu/a = 0.02$ (left) and $\mu/a = 1$ (right). In both cases, as $T\rightarrow 0$, $\bar\kappa\sim T^{c}$, with $c\sim 2$.
}
\label{KScaling}
\end{figure}

Determing the DC thermo-electric conductivities allows us to compare the effects of adding a chemical potential to the thermal conductivity, $\bar \kappa$. For our theory, $\bar \kappa$ is given by
\begin{equation}\label{kappa}
\bar \kappa = \frac{16 \pi^2}{a^2 u_h^3 e^{13\phi_h/4}}T \, .
\end{equation}
Our numerical calculations reveal that the thermal conductivity scales with temperature as $\kappa\sim T^{c}$, where $c$ is a constant with value $\sim 2$ in the case where $\mu = 1$, as shown in figure \ref{KScaling}. We therefore see that the ground state is a thermal insulator. We have checked for various values of $\mu/a$, and the low temperature scaling appears to give the same values.  It is interesting to compare these results to the results from the neutral case \cite{Banks:2015aca}, where it was shown that the neutral black holes have low temperature scaling of $\bar \kappa \sim T^{10/3}$.

\begin{figure}
\centering
\includegraphics[scale = 0.55]{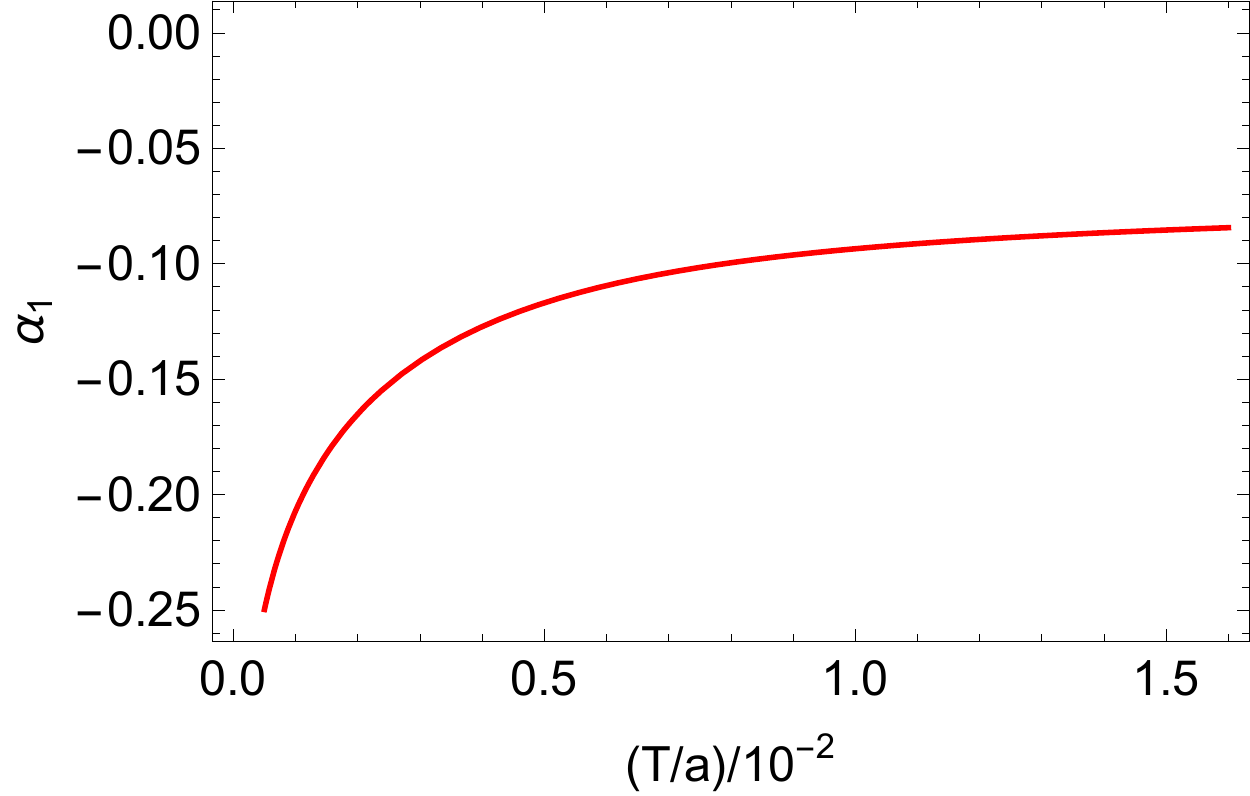} \qquad
\includegraphics[scale = 0.55]{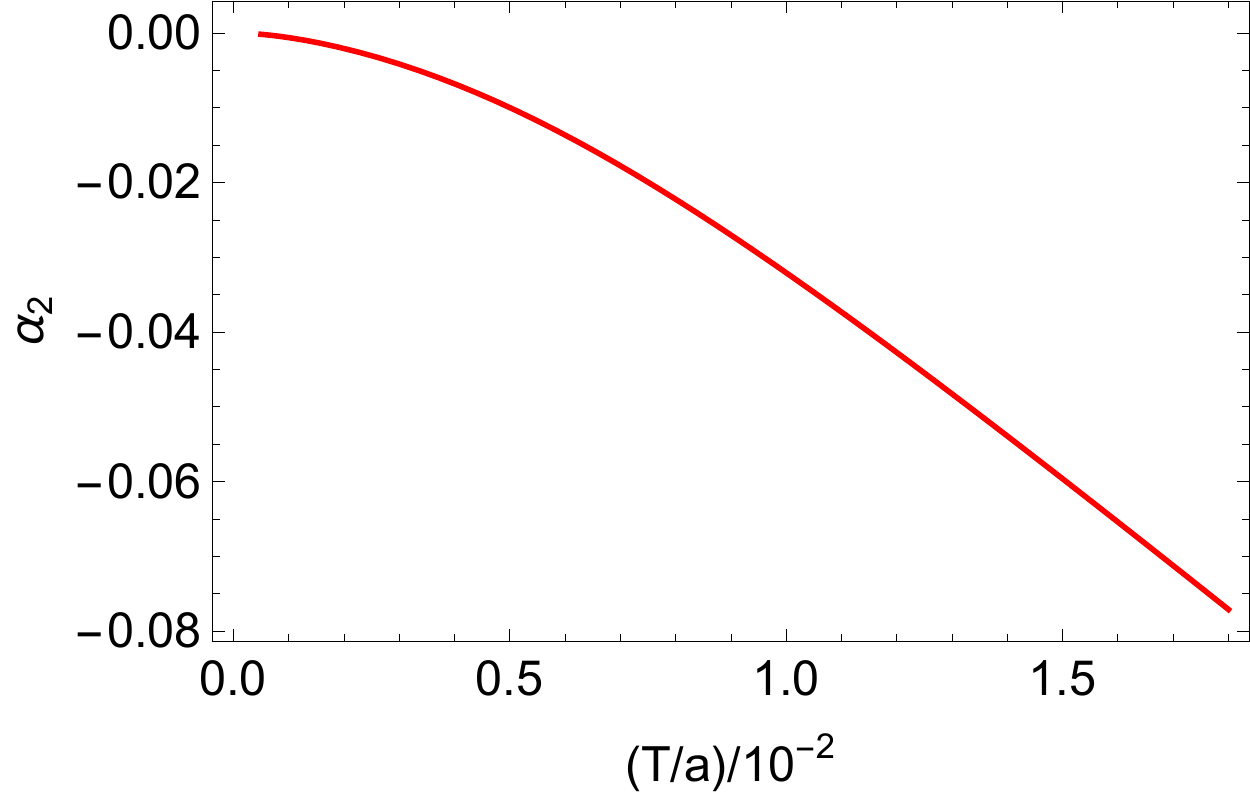}\qquad \,  \\
\includegraphics[scale = 0.55]{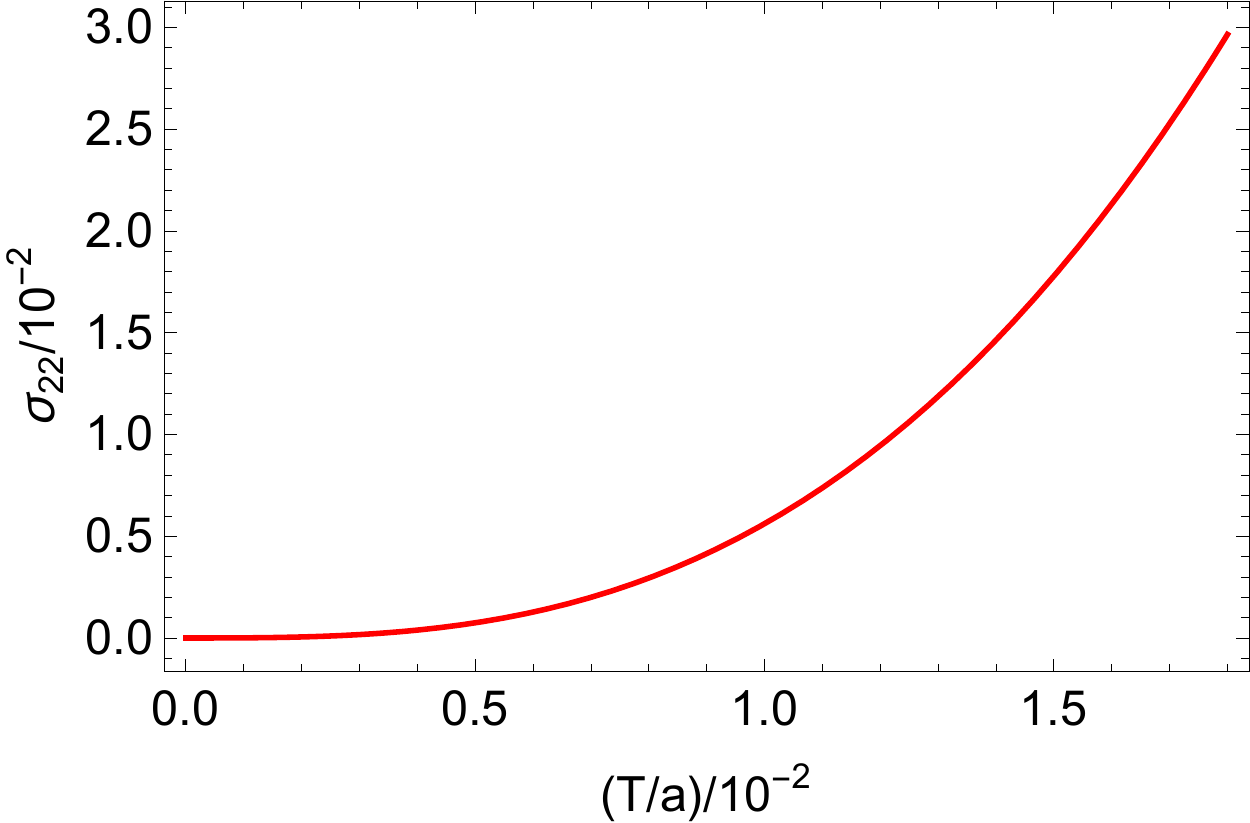}\qquad
\includegraphics[scale = 0.55]{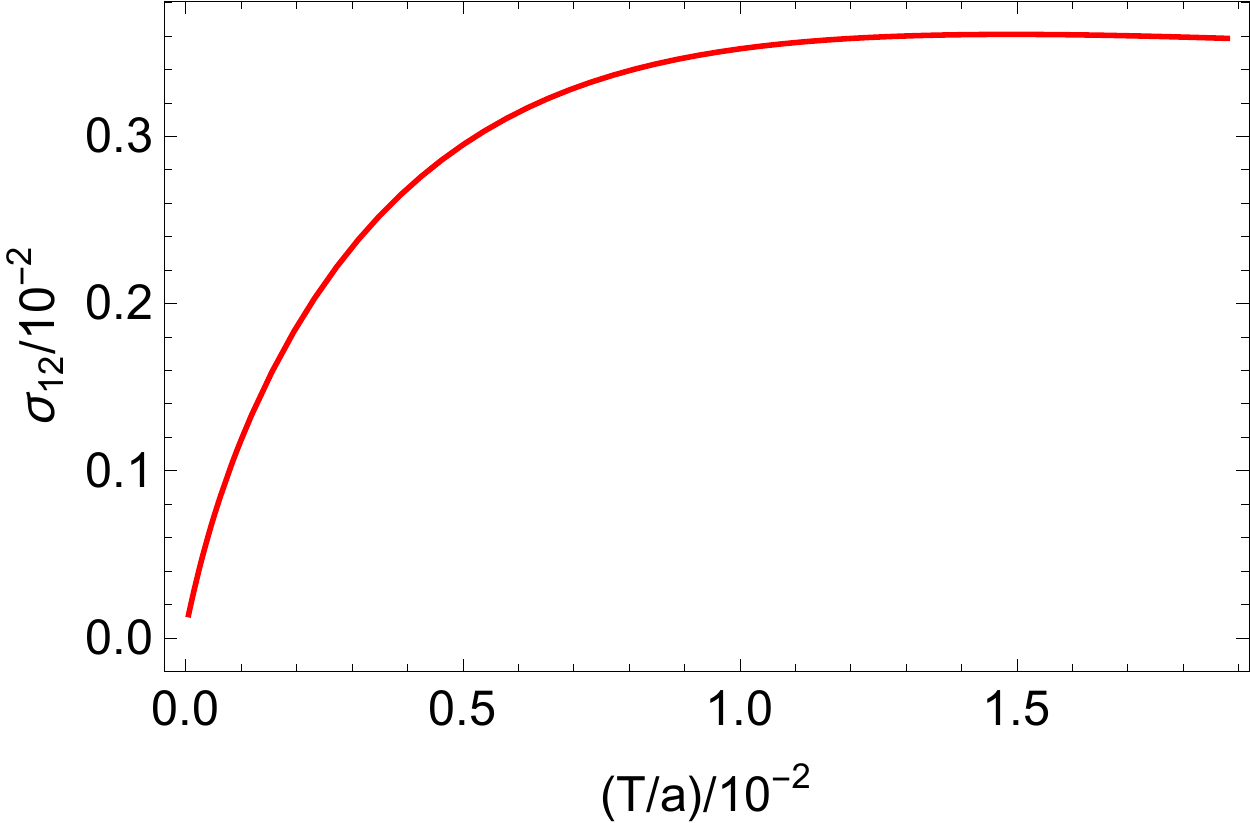}\qquad  \,  \\
\includegraphics[scale = 0.6]{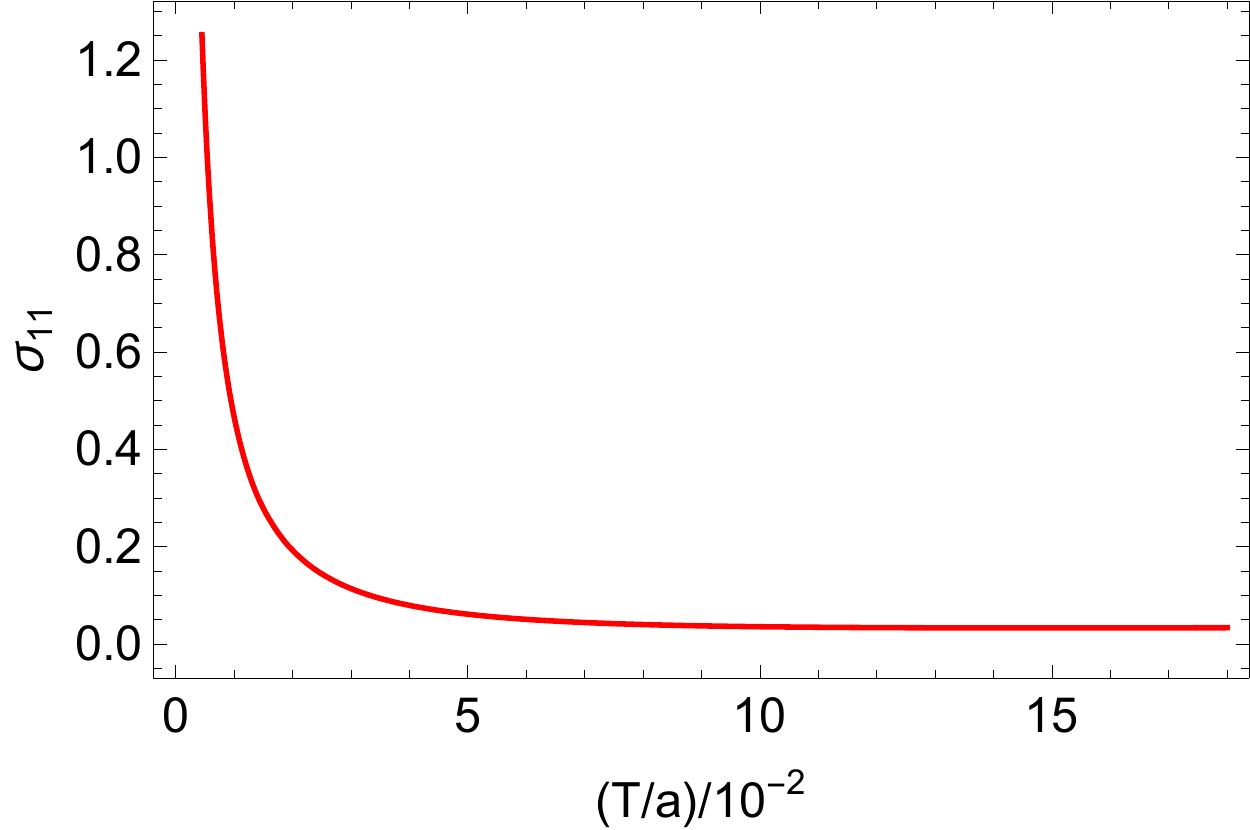}
\caption{Plot of the DC thermoelectric conductivity for the low temperature charged phase, with $\mu/a = 0.02$. We see that both $\sigma_{11}$ and $\alpha_1$ diverge as $T\rightarrow 0$, as the coupling between the $U(1)$ gauge field, $A_1$, and gravity becomes infinite in the bulk theory.}
\label{DCconduc}
\end{figure}

We also find that the electrical conductivties $\sigma_{11}$ and $\alpha_1$ diverge as $T \rightarrow 0$, despite the fact that translational invariance has been explicitly broken in the $z$ direction. To understand this, we note that at as $T\rightarrow 0$, $\psi_{1h}\rightarrow \infty$. Since the coupling of $A_1$ in the action is $\exp{2\psi/\sqrt{6}}$, as $T\rightarrow 0$ this coupling will diverge on the black hole horizon. Therefore, any perturbation of a electric field should lead to an infinite response from the system, and hence an infinite conductivity. Diverging conductivity as $T\rightarrow 0$ has been seen before in the case of AdS-Reisser-Nordstrom black holes, but here it is the coupling, rather than the geometry which is driving the instability. A similar result to this was seen in \cite{Donos:2014uba}.

\section{Discussion}

We have extended the analysis of  \cite{Banks:2015aca}, and constructed a more complete class of black hole solutions that are consistent truncations of the Kaluza-Klein reduction of type IIB supergravity on $S^5$ \cite{Cvetic:2000nc}, and dual to an anisotropically deformed $N=4$ Yang-Mills plasma. We have shown that the low temperature phase from \cite{Banks:2015aca} is actually unstable below a critical temperature, corresponding to an unstable black hole . Whilst the phase transition from \cite{Banks:2015aca} is also present when a finite $U(1)$ chemical potential, $\mu$, is switched on, at $\mu > \mu_c \sim 0.015$, this instability vanishes, and the plasma appears to be stable right down to zero temperature. This charged low temperature plasma has interesting properties, including a divergent electrical conductivity despite the explicit breaking of translational invariance, similar to examples considered in \cite{Donos:2014uba}.

We first considered a consistent truncation that retained five, rather than one, of the 20 scalar fields that transform in the \textbf{20}$'$ of $SO(6)$. At low temperatures, the prescence of additional scalar fields in the truncation gives rise to further  normalisable static modes in the black hole solution constructed in \cite{Banks:2015aca}, and hence new branches of black hole solutions from a critical point $T_{c2}$. Rather than undergoing a phase transition to a thermodynamically preferred phase below $T_{c2}$, however, these new branches of black hole solution only exist at $T> T_{c2}$, and so is an example of retrograde condensation, which has been seen in other top-down holographic models \cite{Aprile:2011uq,Donos:2011ut}. By analysing perturbations of this scalar mode, we can conclude that the solution is unstable below this critical temperature. It would be interesting to understand more generally when retrograde condensation occurs in a top-down setting.

Following this analysis, we turned on a finite chemical potential, $\mu$, in the dual field theory, using a consistent truncation that contains two scalar fields and three $U(1)$ gauge fields. The phase transition first seen in \cite{Banks:2015aca} is also present in the case with a finite chemical potential, with the same order and critical exponents in both cases. In this case, the black hole in \cite{Cheng:2014qia} undergoes a phase transition to a new charged branch of solution. Interestingly, whilst the black hole in  \cite{Cheng:2014qia} reaches zero temperature at some finite radius - an extremal black hole - the new branch of solution does is not extremal at $T = 0$. Whilst the phase transition from the black hole studied in \cite{Cheng:2014qia} occurs at all values of the chemical potential, once $\mu > \mu_c \sim 0.015$, the lower temperature instability discussed above is no longer seen - instead, the solution appears to be stable right down to zero temperature solutions.

However, these two truncations are not the end of the story. As with any top-down model, one is always free to include more of the scalar and gauge fields to analyse the full theory. It would be particularly interesting to understand whether the supergravity solutions of \cite{Mateos:2011ix,Mateos:2011tv,Cheng:2014qia} are truly unstable at low temperatures, or if the full matter content leads to a stable phase. It would also be interesting to understand why the presence of a chemical potential removes this instability. To answer both of these questions would require the full consistent truncation from \cite{Cvetic:2000nc}, which preserves 20 scalar fields and 15 gauge fields. Whilst a technically challenging task, this would allow us to further understand the phase diagram of the anisotropically deformed $N=4$ Yang-Mills plasma. However, in order to fully understand the phase diagram, one would have to construct all the solutions of the dual supergravity theory, not just those with a consistent truncation, and so one would ultimately construct black hole solutions in the full $D = 10$  theory.

Furthermore, it can be shown that the $D = 5$ model preserving the metric, axion, dilaton as well as a single $U(1)$ gauge field can come from a consistent truncation on any five-dimensional Sasaki-Einstein (SE) manifold, not just the five-sphere \cite{Buchel:2006gb}. Therefore, the CGS and Mateos-Trancanelli solutions describe the high temperature phase of a whole class of dual $\mathcal{N} = 1$ spatially anisotropic plasmas. For a consistent truncation on an arbitrary SE manifold,  the unstable scalar modes desribed here may not exist, and so the plasma would have a different phase diagram. It would be interesting to understand precisely how the results here change for different SE manifolds - since a truncation on an arbitrary SE manifold will not have the same scalar field structure, one may have to construct the black holes directly in ten dimensions.

We studied the thermo-electric response of the new branch of black hole solutions at finite chemical potential. We numerically determined the thermal conductivity, $\bar \kappa$, and showed that the plasma is a thermal insulator. Furthermore, we determined the $\bar \kappa$ scales at low temperatures by $T^c$, with $c\sim 2$ in the case where $\mu/a = 1$. In \cite{Banks:2015wha}, the authors were able to construct the IR behaviour of the zero temperature black hole solution, and found it obeyed a scaling relation with a metric that was both Liftshitz and hyperscaling violating. It would be an interesting avenue of further work to try to find a similar solution in this case. This could possibly shed light on the reason for why a chemical potential stabilises the low temperature theory. 

At low temperatures, part of the DC thermo-electric conductivity matrix diverges, despite the explicit breaking of translationl invariance in the model. This is because the gauge field coupling in the theory diverges as $T \rightarrow 0$, and so any perturbation of the electric field will give an infinite response. This highlights a further challenge in modelling physical systems in a top down setting. In order to introduce finite DC conductivity in a holographic setting, one has to break translationa invariance, such as through a spatially dependent source term. However, many top down models contain non-linear couplings between gauge and scalar fields, and so these couplings may well diverge at certain temperatures. Therefore, even when there is a mechanism for momentum dissipation, top down models can still have infinite DC thermo-electric conductivities.

All of the previous works have focused on case of static black hole solutions, which correspond to electrically charged field theories. However, it would be interesting to understand how this system behaves when there is also a magnetic field present. In this case, the black holes would no longer be static, but simply stationary, and the Chern-Simons term from (\ref{QLagrangian}) would no longer be zero. Whilst this would be technically challenging, a starting point could be to construct black hole solutions similar to \cite{Cheng:2014qia} that contains magnetic fields.

\section*{Acknowledgements}
EB would like to thank Jerome Gauntlett, Toby Wiseman and Aristomenis Donos for helpful discussions. EB would also like to thank Sang-Jin Sin, Xian-Hui Ge and Long Cheng for useful explanations of the methods used in their work. The work is supported by an Imperial College Schrodinger Scholarship.

\appendix
\renewcommand{\thesection}{\Alph{section}}

\section{Zero temperature charged black hole solutions} \label{Extremal}

\subsection{Extremal black hole}

\begin{figure}
\centering
\includegraphics[scale = 0.55]{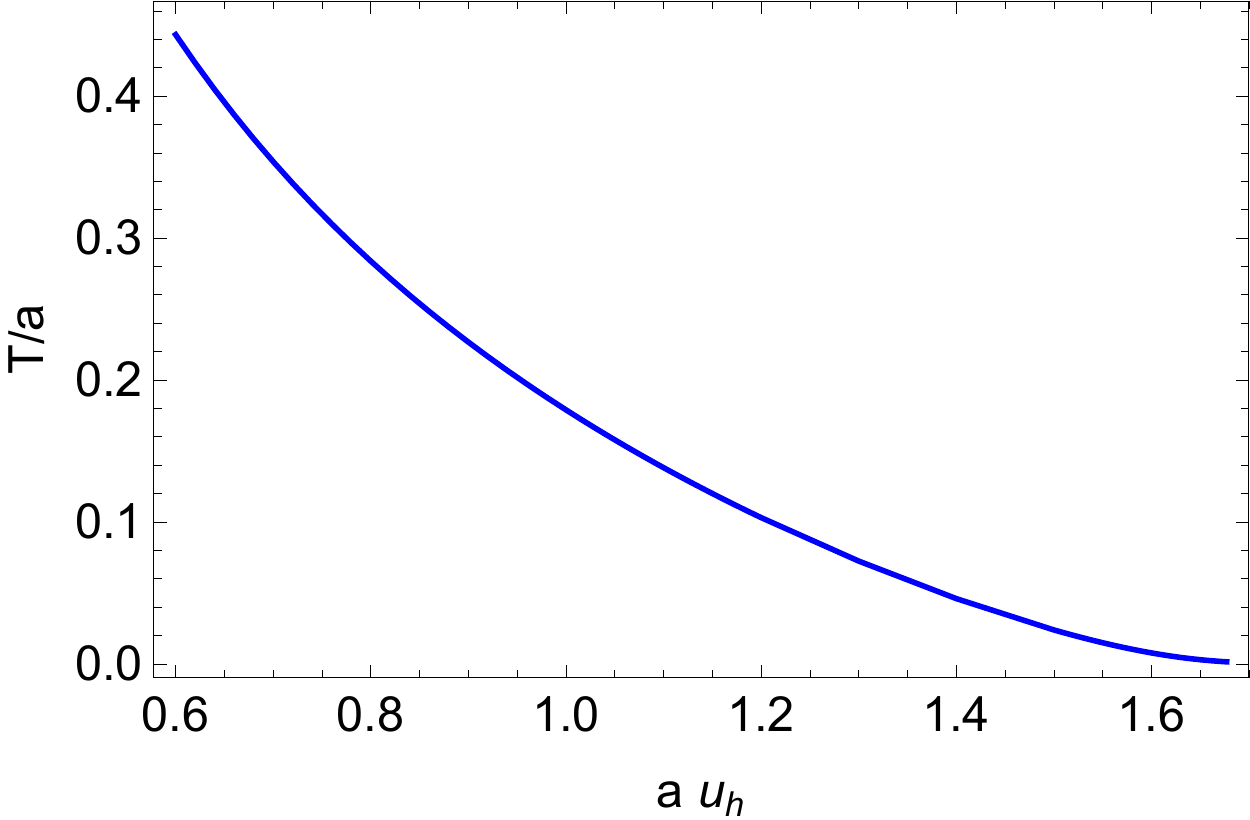} \qquad
\includegraphics[scale = 0.55]{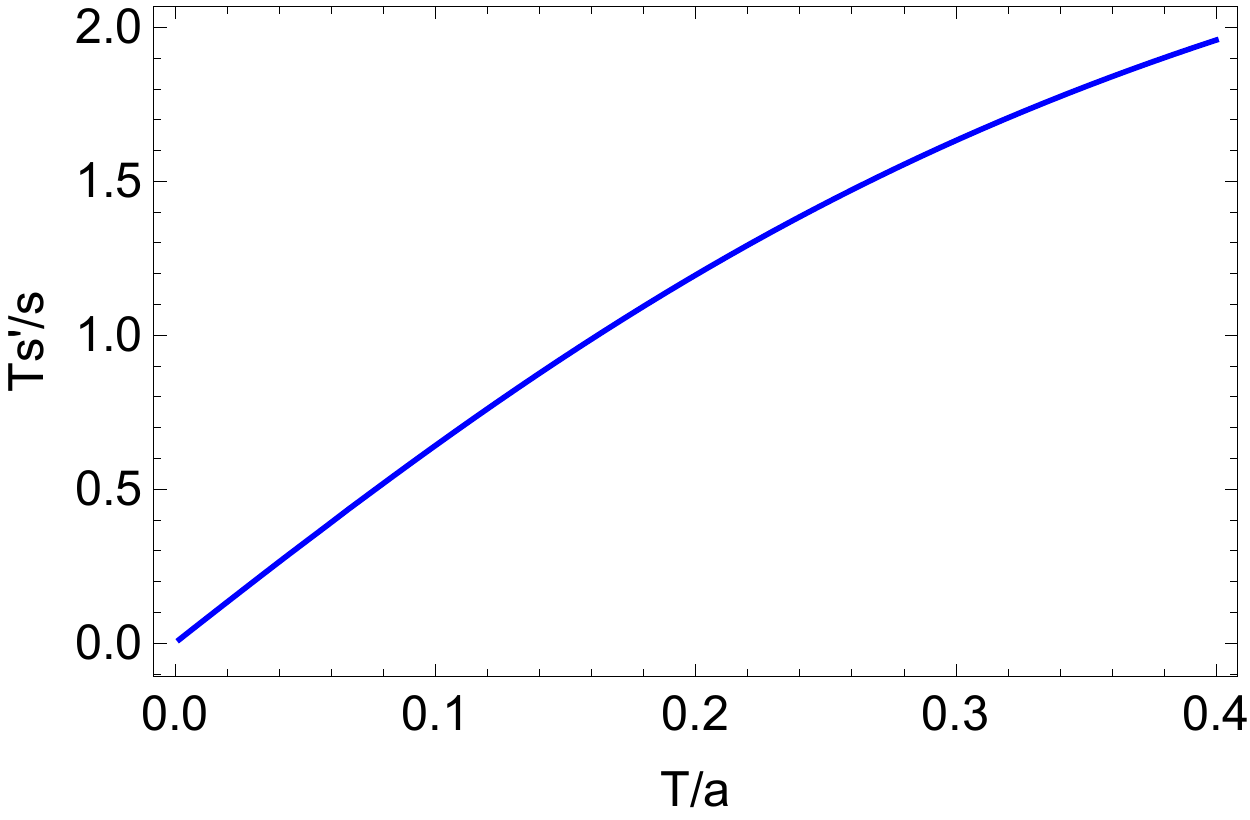} 
\caption{Plot showing charged CGS solution forming an extremal black hole, when $\mu/a = 1$. The right plot shows how $T/a$  changes with $u_h a$. As we as we can tell, there is a one to one mapping, and hence there is no Hawking-Page transition. The left plot shows the temperature scaling of entropy. As $T/a \rightarrow 0$, the entropy tends to a constant value, and hence the black hole is extremal.}
\label{Extremal Black Hole}
\end{figure}

We will now discuss some low temperature features of the CGS solution from \cite{Cheng:2014qia}. It is interesting to note that we did not observe the Hawking-Page transition that the authors claimed appeared in their work. There, it was argued that there are different black hole radii corresponding to the same temperature, and hence there is a minimum black hole temperature and the solution is unstable below this temperature. However, we have not found this. Rather, for a fixed chemical potential, one can cool the solution down to arbitrarily low temperatures, as shown in the left hand plot of figure \ref{Extremal Black Hole}. We note that this is a different to the results of \cite{Iizuka:2012wt}, where a model containg a linear axion, a dilaton and U(1) gauge field was studied, and no extremal horizon was found. However, in that case the gauge field coupled to the dilaton, whereas here there is no coupling between the gauge field and dilaton.

Futhermore, the zero temperature limit of the CGS solution is at a finite black hole radius, and so the black hole is an extremal black hole. As shown in the right plot of figure \ref{Extremal Black Hole}, the leading order scaling is $s \sim T^0$, and hence the entropy of the black hole tends to a constant value in the zero temperature limit. In addition, the functions $\mathcal{F},\mathcal{F}',\mathcal{B},\mathcal{B}'$ all vanish on the horizon, indicating a second order pole in the metric on the horizon and an extermal black hole. Whilst this is worrying from a physical perspective, as discussed in the main text, the black hole is unstable and therefore this extremal black hole would never be realised in reality.

\subsection{Weak anisotropic limit}
We now wish to study this zero temperature extremal black hole in the limit where $\mu/a >> 1$, i.e the limit of weak anisotropy. At zero temperature and $a = 0$, our solution is just the extremal AdS-Reissner-Nordstrom black hole, and we therefore look for an expansion of the form
\begin{align}
\mathcal{F} &= \left(1 - \frac{(\mu u)^4}{12} + \frac{(\mu u)^6}{108}\right) + a^2\mathcal{F}_2(u) + \mathcal{O}(a^4) \, ,\nn
\mathcal{B} &= 1 + a^2\mathcal{B}_2(u) + \mathcal{O}(a^4) \, ,\nn
\phi &= a^2\phi_2(u) + \mathcal{O}(a^4) \, ,\nn
b &= \mu\left(1 -\frac{(\mu u)^2}{6}\right) +  a^2b_2(u) + \mathcal{O}(a^4) \, ,
\end{align}
where $\mu$ from (\ref{uvexp}) has been rescalled by $\mu \rightarrow \mu/\sqrt{3}$, as per the discussion in section (\ref{Charged model}). Note that only even powers of $a$ are allowed, due to the symmetry $z \rightarrow -z$. In addition, the location of the extremal horizon will also have an expansion, given by
\begin{equation}
u_h = \frac{\sqrt{6}}{\mu} + a^2 u_h^{(2)} +  \mathcal{O}(a^4) \, .
\end{equation}

\begin{figure}
\centering
\includegraphics[scale = 0.7]{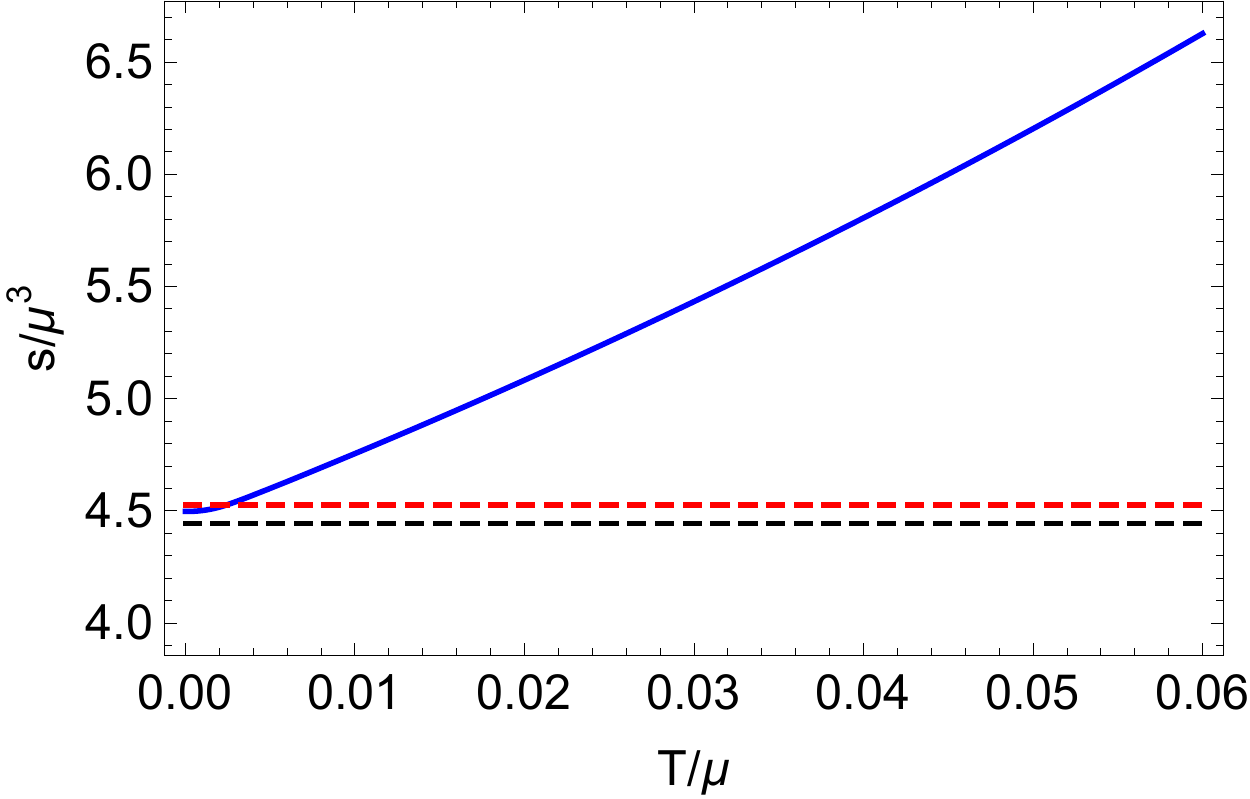}
\caption{Plot showing charged enrtropy density versus temperature, when $\mu = \sqrt{3}$ and $a = 1/10$. The blue line is the entropy density, whilst the red dashed line is the analytical expression at zero temperature for $s$ to $\mathcal{O}(a^2)$, and the black line is the entropy of the extremal AdS-Reissner-Nordstrom black hole. The result is in good agreement with the analytical expression.}
\label{EntPlots}
\end{figure}

Now we substitute these expansions into the equations of motion, and solve order by order in $a$. We impose the boundary conditions that all the terms of $\mathcal{O}(a^2)$ and higher vanish at the UV boundary, which ensures the solution approaches $AdS_5$ in the UV. We also require $b$ and $\mathcal{F}$ to vanish on the horizon at each order in $a$. We fix the $u^{(i)}_h$ by setting the temperature of the black hole to be zero at each order of $a$. After solving the equations of motion, we find analytical solutions at leading order
\begin{align}
\mathcal{B}_2(u) &= \frac{90 \mu ^2 u^2+\left(\mu ^2 u^2-6\right) \left(\mu ^2 u^2+3\right) \log \left(6-\mu ^2 u^2\right)+\left(3 \mu ^2 u^2+18-\mu ^4 u^4\right) \log \left(2 \mu ^2 u^2+6\right)}{12 \mu ^2 \left(\mu ^2
   u^2-6\right) \left(\mu ^2 u^2+3\right)} \, , \nn
\mathcal{F}_2(u) & = \frac{u^6\mu^4 \left(20+24 \tanh ^{-1}\left(\frac{1}{9} \left(3-2 \mu ^2 u^2\right)\right)-2\left(5 \log \left(\frac{5 a^2}{9 \mu ^2}\right)+\log 2\right)\right)}{1296}+ \, \nn & +  \frac{u^4\mu^2
   \left(60 \log \left(\frac{5a^2}{\mu ^2}\right)-126 \tanh ^{-1}\left(\frac{1}{9} \left(3-2 \mu ^2 u^2\right)\right)-192+ \log 8-120  \log 3\right)}{1296} \, \nn 
&+\frac{ \log \left(\frac{18+6\mu ^2 u^2}{18-3 \mu ^2 u^2}\right)}{4 \mu ^2}+\frac{u^2}{3} \, , \nn
b_2(u)&= \frac{5}{72} \mu  u^2 \left(\log \left(\frac{5a^2}{\mu ^2}\right)+\log \left(\frac{\mu ^2 u^2+3}{54-9 \mu ^2 u^2}\right)-2\right) \, , \nn
\phi_2(u) &= \frac{1}{2 \mu ^2}\log \left(\frac{6-\mu ^2 u^2}{6+2\mu ^2 u^2}\right) \, ,
\end{align}
with $u^{(2)}_h = -5/(2\sqrt{6}\mu^3)$. Using these expressions, we can then get an expansion for the entropy density of the extremal black hole
\begin{equation}
\frac{s}{\mu^3} =  \sqrt{\frac{2}{3}}\frac{ \pi }{3} +\frac{5 \left(2-\log \left(\frac{5a^2}{18 \mu ^2}\right)\right)}{48 \sqrt{6}}\left(\frac{a}{\mu}\right)^2 + \mathcal{O}\left(\frac{a}{\mu}\right)^4 \, .
\end{equation}
This gives us the leading correction to the zero temperature entropy density, in the weak anisotropic limit. In figure \ref{EntPlots}, we plot entropy against temperature, and see that for $a/\mu << 1$, the entropy in the zero temperature limit is consistent with the analytic expansion. This is further evidence that the low temperature CGS solution is extremal and does not undergo a Hawking-Page type transition.

\subsection{Fixed point solution}

To understand zero temperature black holes, important physics can often come from scaling solutions. To understand the extremal CGS solution for arbitrary $a/\mu$, we will therefore look for scaling solutions to the CGS equations of motion of the form
\begin{gather}
e^{\phi} = e^{\phi_0}u^{\phi_c}, \, \, \,  \mathcal{F} = \mathcal{F}_0u^{\mathcal{F}_c}, \,\,\, \mathcal{B} = \mathcal{B}_0u^{\mathcal{B}_c}, \,\,\, b = b_0u^{b_c} \, ,
\end{gather}
where $b$ is the single gauge field that we have in the CGS solution, and recall that there are no additional scalar fields in the CGS solution.

Remarkably, we find that there is an exact solution to the equations of motion
\begin{gather}
e^{\phi} = \phi_0 (au)^{12/5}, \, \, \,  \mathcal{F} =\frac{25}{192}\phi_0^3 (au)^{26/5}, \,\,\, \mathcal{B} = \frac{32}{75}\left(\frac{\mu}{a}\right)^2 \phi_0(au)^{6}, \,\,\, b = \mu (au)^{16/5} \, ,
\end{gather}
where $\mu$ and $\phi_0$ are constants. After writing $u = c\rho^{5/16}$ for some constant, $c$, this solution can be written as
\begin{align}
ds^2 \sim & \rho^{\frac{-2(3-\theta)}{3}}\left(d\rho^2-d\bar t^2+\rho^{-2(z_1-1)}\left(d\bar x^2+d\bar y^2\right)+\rho^{-2(z_2-1)}d\bar z^2\right)\,,\nn
&e^\phi\sim\rho^{3/16},\qquad a\sim \rho\,,
\end{align}
with $\theta = 3$, $z_1 = 9/8$, $z_2 = 3/4$ and the bars indicate we have rescaled the coordinates. This is reminiscent of hyper-scaling solutions with hyper-scaling violation exponent $\theta$, but here the spatial directions also scale with Liftshitz exponents $z_1$ and $z_2$. Under the scaling $(\bar t, \bar x, \bar y, \bar z, \bar \rho) \rightarrow (\lambda \, \bar t,\lambda^{9/8} \bar x,\lambda^{9/8} \bar y,\lambda^{3/4} \bar z, \bar \rho)$, we find that the metric transforms as $ds \rightarrow \lambda^{\theta/3} ds$.

Interestingly, the determinant of this metric does not depend on the radial coordinate. Therefore, if this solution can be generalised to a finite temperature black hole, then the area of this black hole would be the same at any radius and so the solution would have constant entropy. This might suggest that this scaling solution is related to the extremal black hole solution discussed above.

To see whether this solution can be heated to finite temperature, we consider static perturbations about the fixed point solution
\begin{align}
e^{\phi} &= e^{\phi_0}u^{\phi_c}(1+c_1 u^{\delta}), \, \qquad  \mathcal{F} = \mathcal{F}_0u^{\mathcal{F}_c}(1+c_2 u^{\delta}), \,\,\, \nn
 \mathcal{B} &= \mathcal{B}_0u^{\mathcal{B}_c}(1+c_3 u^{\delta}), \,\, \qquad b = b_0u^{b_c}(1+c_4 u^{\delta}) \, .
\end{align}
Substituting this ansatz into the equations of motion, and keeping terms linear in $c_i$, we find three solutions. Two of these are marginal modes with $\delta = 0$, and correpond to the scaling symmetries (\ref{scsym}), whilst the third corresponds to the gauge transformation $b \rightarrow b + c$, for some constant c. This suggests that there are insufficient parameters to develop an IR expansion, and hence create a solution that interpolates between this scaling solution in the IR and $AdS_5$ in the UV.

Therefore, it appears that this scaling solution is not the fixed point solution in the extremal limit of the CGS black hole. It would be an interesting topic of futher work to understand the extremal CGS black hole for arbitrary charge, as it may shed light on the mechanism for the low temperature instability of the CGS black hole.

\section{Thermo-electric DC conductivity with multiple gauge fields} \label{DC}

We will now derive the DC thermo-electric conductivity as outlined in section \ref{Charged}. To start with, we will consider a more general theory, and derive its general conductivity matrix. Specifically, we generalise \cite{Banks:2015wha} to the case where there are multiple $U(1)$ gauge fields, as well as additional scalars. Using the same notation as \cite{Donos:2015gia,Banks:2015wha}, we generalise the Lagrangian from \cite{Banks:2015wha} so that we consider a general Lagrangian
\begin{align}\label{eq:bulk_action}
S=\int d^D x \sqrt{-g}\,\left(R-V(\phi)-\sum_a\frac{Z_a(\phi)}{4}\,(F^a)^{2}-\frac{1}{2}\mathcal{G}_{IJ}\partial\phi^I \partial\phi^J \right)\,.
\end{align}
where $a$ is a finite number of $U(1)$ gauge fields, and the only restriction on each of the $Z_a$ is that $Z_a(0)$ is constant. We consider black hole solutions of the form
 \begin{align}\label{eq:DC_ansatz}
ds^{2}&=-UG\,dt^{2}+\frac{F}{U}\,dr^{2}+ds^2(\Sigma_d)\,, \nn
A^a&=a^a_{t}\,dt\,,
\end{align}
where
$ds^2(\Sigma_d)\equiv g_{ij}(r,x)dx^i dx^j$ is a metric on a ($d\equiv D-2$)-dimensional manifold, $\Sigma_d$,
at fixed $r$. In addition, $U=U(r)$, while $G,F,a_t$ and $\phi$ are all functions of 
$(r,x^i)$. 

The boundary conditions are chosen to ensure that the solution approaches $AdS_D$ as $r \rightarrow \infty$. These are the same black hole solutions as in \cite{Banks:2015wha}. We now perturb this solution with a linear perturbation
\begin{align}\label{pertansatz}
\delta \left(ds^2\right)&=\delta g_{\mu\nu} dx^\mu dx^\nu-2t M\zeta _i dt dx^i\,,\nn
\delta A^a&=\delta a^a_\mu dx^\mu-t E^a_i dx^i+t N^a \zeta_i dx^i\,,\nn
\delta\phi^I\,.
\end{align}

The calculation of the DC conductivity now proceeds in very similar way to \cite{Banks:2015wha}. Rather than repeat the entire calculation, we just quote the final important results, keeping the same notation as in the original paper. The Hamiltonian constraints evaluated on the black hole horizon lead to the Stokes equations, which are now
\begin{align}
\nabla_{i} v^{i}=0\,,& \label{eq:v_eq2}\\
\nabla_i(Z_a^{(0)}\nabla^i w^a)+v^{i}\,\nabla_{i}\left(Z_a^{(0)}{a^a}_{t}^{(0)} \right)=-\nabla_{i}(Z_a^{(0)} {E^a}^{i})\,,&\label{eq:w_eq2}\\
-2\,\nabla^{i}\nabla_{\left( i \right. }v_{\left. j\right)}-\sum_a\frac{Z_a^{(0)}{a^a}_{t}^{(0)}}{G^{(0)}}\nabla_j w^a
+{\cal G}_{IJ}(\phi^{(0)})\nabla_j\phi^{I(0)}\nabla_i\phi^{J(0)}v^{i}& \nn 
+\nabla_{j}\,p=4\pi T\,\zeta_{j}
+\sum_a \frac{Z_a^{(0)}{a^a}_{t}^{(0)}}{G^{(0)}} E^a_j\,,\label{eq:V_neutral2}
\end{align}
where the $a$ indicies are only summed over explicity, and 
\begin{align}\label{deffquants}
v_{i}\equiv -\delta g_{it}^{(0)},\qquad w^a\equiv \delta {a^a}_{t}^{(0)},\qquad p\equiv -4 \pi T\frac{\delta g_{rt}^{(0)}}{G^{(0)}}-\delta g_{it}^{(0)}g^{ij}_{(0)}\nabla_{j}\,\ln G^{(0)}\,,
\end{align}
where the $(0)$ indicates the leading order term in the expansion of the field about the horizon, and $F^{(0)} = G^{(0)}$.

The heat current and electric currents are given by
\begin{align}\label{eq:J_hor2}
Q^i_{(0)} &=4\pi T\sqrt{g_{(0)}}v^j\,,\nn
{J^a}^i_{(0)}&=\sqrt{g_{(0)}}g^{ij}_{(0)}Z_a^{(0)}\left(\partial_j w^a+\frac{{a^a}^{(0)}_t}{G^{(0)}}v_j+E^a_j\right)\,.
\end{align}

We now explicitly consider black hole solutions where there are scalars associated with a shift symmetry. In our explicit example that is the axion field. In general, these scalar fields take the form
\begin{align}
\phi^{I_{\alpha}}=\mathcal{C}^{I_{\alpha}}{}_{j}\,x^{j}\,,
\end{align}
everywhere in bulk with $\mathcal{C}$ a constant $n$ by $d$ matrix. For simplicity in this general case, we assume that all spatial coordinates are involved and hence the DC conductivity in all spatial
directions is finite. The metric, the gauge fields and the remaining scalar fields will depend on the radial direction but will be independent of the spatial coordinates $x^i$. 
The metric on the black hole horizon is flat and in addition, $Z^{(0)}$, $G^{(0)}$ and $a^{(0)}_t$ are all constant.

There is a solution to the fluid equations \eqref{eq:v_eq2}-\eqref{eq:V_neutral2}, with $v^{i}$, $p$ and $w$ all constant on the horizon. The fluid velocity is given by
\begin{align}
v^{i}&=4\pi T\,\left(\mathcal{D}^{-1} \right)^{ij}\,\left(\zeta_{j}+\frac{1}{Ts}\,\sum_a \rho_aE^a_{j} \right)\,,
\end{align}
with constant $E_i,\zeta_i$ and we have defined the $d\times d$ matrix:
\begin{align}
\mathcal{D}_{ij}&=G_{I_{\alpha_{1}} I_{\alpha_{2}}}\,\mathcal{C}^{I_{\alpha_{1}}}{}_{i}\,\mathcal{C}^{I_{\alpha_{2}}}{}_{j}\,.
\end{align}
Furthermore, the averaged charge density, $\rho$,
and the entropy density, $s$, are given by
\begin{align}
\rho_a=\sqrt{g_{(0)}}\frac{Z_a^{(0)}{a^a}_{t}^{(0)}}{G^{(0)}},\qquad s=4\pi \sqrt{g_{(0)}}\,.
\end{align}
The current densities $J^i,Q^i$ are independent of the radius and are given by their
horizon values:
\begin{align}
{J^a}^{i}=& \frac{s\,Z_a^{(0)}}{4\pi}\,g_{(0)}^{ij}\,E^a_{j}+ \frac{4\pi\rho_a}{s}\,\sum_b\rho_b\left( \mathcal{D}^{-1}\right)^{ij} \,E^b_{j}+4\pi T\rho_a \left( \mathcal{D}^{-1}\right)^{ij}\zeta_{j}\,,\nn
Q^{i}=&4\pi Ts \left( \mathcal{D}^{-1}\right)^{ij}\,\left(\zeta_{j}+\sum_a \rho_aE^a_{j}\right)\,.
\end{align}
The DC conductivities are thus given by
\begin{align}
\sigma^{ij}_{ab}=&\frac{s\,Z_a^{(0)}}{4\pi}g_{(0)}^{ij}\delta_{ab}+ \frac{4\pi\rho_a\rho_b}{s}\,\left( \mathcal{D}^{-1}\right)^{ij}\,,\nn
\alpha_a^{ij}=&\bar{\alpha}_a^{ij}=4\pi\rho_a \,\left( \mathcal{D}^{-1}\right)^{ij}\,,\nn
\bar{\kappa}^{ij}=&4\pi T s\,\left( \mathcal{D}^{-1}\right)^{ij}\,.
\end{align}

We now return to the black hole solution in section (\ref{Charged}), which preserves two gauge fields and one scalar field ( in addition to the axion and the dilaton). In this case, the only the $z$ direction will have a finite conductivity matrix. In the notation of \cite{Banks:2015wha}, have $G = F = \frac{e^{-\phi/2}}{u^2}\sqrt{\mathcal{B}}$ and $U = \mathcal{F}\sqrt{\mathcal{B}}$. We also have $Z^{(0)}_1 = Z^{(0)}_2 =e^{2\psi_{1h}/\sqrt{6}}$ and $Z^{(0)}_3 =e^{-4\psi_{1h}/\sqrt{6}}$, which leads to the charge densities
\begin{align}
&\rho_1=\rho_2=\frac{a_{1h}e^{2\psi_{1h}/\sqrt{6}}}{u_h e^{3\phi_h/4}\sqrt{\mathcal{B}_h}},\qquad  \rho_3 = \frac{a_{3h}e^{-4\psi_{1h}/\sqrt{6}}}{u_h e^{3\phi_h/4}\sqrt{\mathcal{B}_h}},
\end{align}
while the entropy density is given as in (\ref{ent}). Finally, the matrix $\mathcal{D}$ is given by
\begin{equation}
\mathcal{D}_{33} = a^2e^{2\phi} \,
\end{equation}
and so the conductivity matrix can now be determined.


\providecommand{\href}[2]{#2}\begingroup\raggedright\endgroup

\end{document}